\documentclass[%
reprint,
superscriptaddress,
 amsmath,amssymb,
 aps,
prd,
]{revtex4-2}

\usepackage{graphicx}
\usepackage{dcolumn}
\usepackage{bm}

\usepackage{xcolor}

\begin{document}


\title{Time delay interferometry with minimal null frequencies}

\author{Gang Wang}
\email[Gang Wang: ]{gwang@shao.ac.cn, gwanggw@gmail.com}
\affiliation{Shanghai Astronomical Observatory, Chinese Academy of Sciences, Shanghai 200030, China}

\date{\today}

\begin{abstract}

Time delay interferometry (TDI) is a key technique employed in gravitational wave (GW) space missions to mitigate laser frequency noise by combining multiple laser links and establishing an equivalent equal arm interferometry. The null frequencies will be introduced in noise spectra and GW response when the periodical signal/noise is canceled in synthesized laser links. These frequencies are characteristic frequencies (CFs) of a TDI which related to its geometry of combination. 
In this work, we implement a second-generation TDI configuration referred to as hybrid Relay to perform noise suppressions and data analysis, whose CFs are only one-quarter that of the fiducial second-generation Michelson observables. We examine the performance of TDI configuration in laser noise cancellation and clock noise suppression and justify its essential capabilities. To assess its robustness for signal extraction, we simulate data containing GW signals from massive black hole binaries and perform parameter inferences with comparisons against the fiducial Michelson TDI configuration. The results demonstrate that the alternative TDI solution could be more robust than Michelson in fulfilling data analysis.

\end{abstract}


\maketitle

\section{Introduction}

LISA is scheduled to be launched in the mid-2030s with a mission to observe gravitational waves (GWs) in the milli-Hz band \cite{2017arXiv170200786A,Colpi:2024xhw}. The detector comprises three spacecraft (S/C) forming a triangular constellation, utilizing laser interferometry between S/C to monitor the distance fluctuations caused by GWs. However, due to orbital dynamics and laser stability limitations, laser frequency noise may dominate GW signal in inter-S/C measurements. 

To address this challenge and achieve the target sensitivity, time-delay interferometry (TDI) has been developed, synthesizing multiple laser links to form an equivalent equal-arm interferometer \cite{1999ApJ...527..814A}.
Various TDI configurations have been devised with different abilities to suppress laser noise. The first-generation TDI cancels laser noise in a static unequal arm constellation \cite[and references therein]{1999ApJ...527..814A,2000PhRvD..62d2002E,2001CQGra..18.4059A,Larson:2002xr,Dhurandhar:2002zcl,2003PhRvD..67l2003T,Vallisneri:2004bn,2008PhRvD..77b3002P,Tinto:2020fcc}, while the second-generation TDI further suppresses laser noise in time-varying arm lengths \cite[and references therein]{Shaddock:2003dj,Cornish:2003tz,Tinto:2003vj,Dhurandhar:2010pd,2019PhRvD..99h4023B,2020arXiv200111221M,Tinto:2022zmf}. Modified TDI formulations driven by data analysis have been developed to comprehensively resolve laser noise and/or signals \cite{Vallisneri:2020otf,Page:2021asu,Page:2023hxm,Baghi:2020ygw,Baghi:2022wdd}. Furthermore, advancements in inter-S/C measurement designs have enabled the reduction of clock noise in TDI \cite{Otto:2012dk,Otto:2015,Tinto:2018kij,Hartwig:2020tdu,Hartwig:2021dlc}. 

Current TDI studies primarily focused on the Michelson configuration, which mimics the original Michelson and employs four laser links in two interferometric arms, synthesizing the measurements recurrently. However, the symmetric geometry introduces null frequencies or characteristic frequencies (CFs) where the periodic signals and noises are significantly canceled. Additionally, differences between unequal arms result in non-identical observables from different two arms. These properties of Michelson TDI could raise numerical challenges for data analysis.

In this study, we implement an alternative TDI configuration referred to as hybrid Relay to perform noise suppressions and data analysis. \citet{Wang:2011} first developed this TDI configuration in 2011 and analyzed its average sensitivity in 2020 \cite{Wang:2020pkk}. \citet{Muratore:2020mdf} also identified this configuration in 2020.
One merit of these observables is that they have only one-quarter CFs of second-generation Michelson, resulting in smoother noise spectra and GW response functions.
As a second-generation TDI configuration, the new observables could sufficiently suppress laser frequency noise and clock noise as the Michelson observables.  Another advantage is that the optimal TDI channels from the alternative configuration would be more robust than Michelson, avoiding susceptibility to unequal arm lengths during orbital motion. 
To validate the performance of the proposed TDI configuration, simulation and analysis are conducted using SATDI \cite{Wang:2024a}. With the data containing GW signals from massive binary black holes (MBBH), the hybrid Relay demonstrates a stable capability to yield reliable results. Further investigation shows that the hybrid Relay also exhibits a more robust ability to characterize the instrumental noises compared to the Michelson configuration \cite{Wang:2024hgv}.
These advantages make it a promising TDI candidate for space missions GW data processing. 

This paper is organized as follows:
We introduce the hybrid Relay TDI configuration and its geometry in Section \ref{sec:tdi}. 
In Section \ref{sec:sensitivity}, the capabilities of laser noise suppression and clock noise cancellation are examined for proposed TDI observables, and the averaged sensitivities are evaluated for the ordinary and optimal channels.
The data simulation and GW analysis are presented in Section \ref{sec:simulation_analysis}. Post-analysis further reveals the disadvantages of the Michelson configuration and the advantages of the hybrid Relay.
A brief conclusion and discussion are given in Section \ref{sec:conclusions}.
(We set $G=c=1$ in this work except specified in the equations.)

\section{Time delay interferometry of hybrid Relay} \label{sec:tdi}

\begin{figure}[htb]
\includegraphics[width=0.18\textwidth]{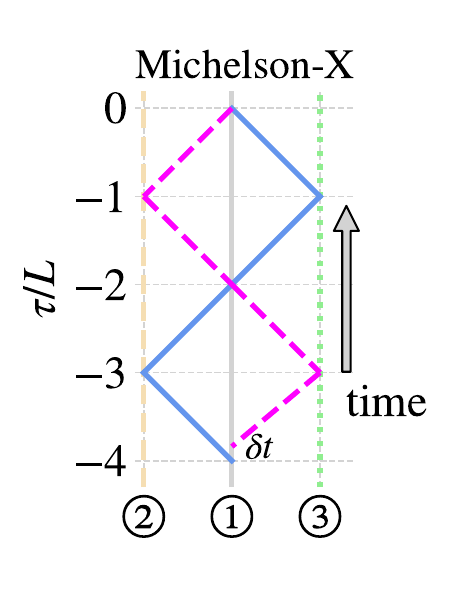}
\includegraphics[width=0.27\textwidth]{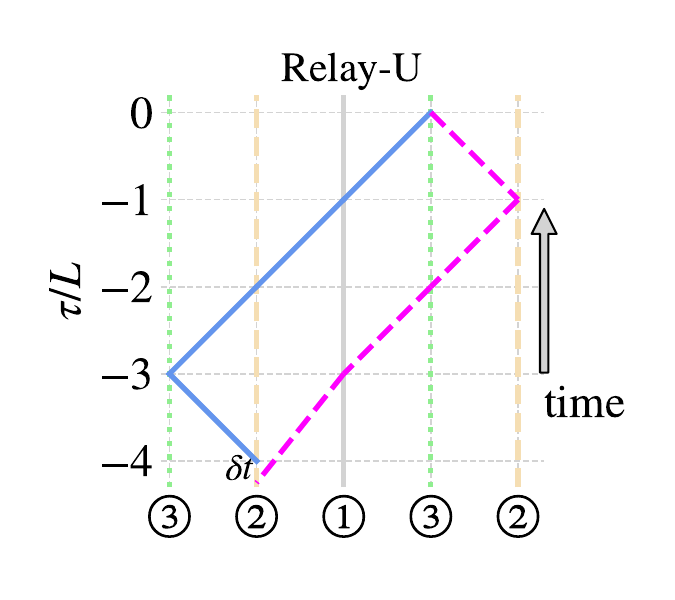} 
\includegraphics[width=0.27\textwidth]{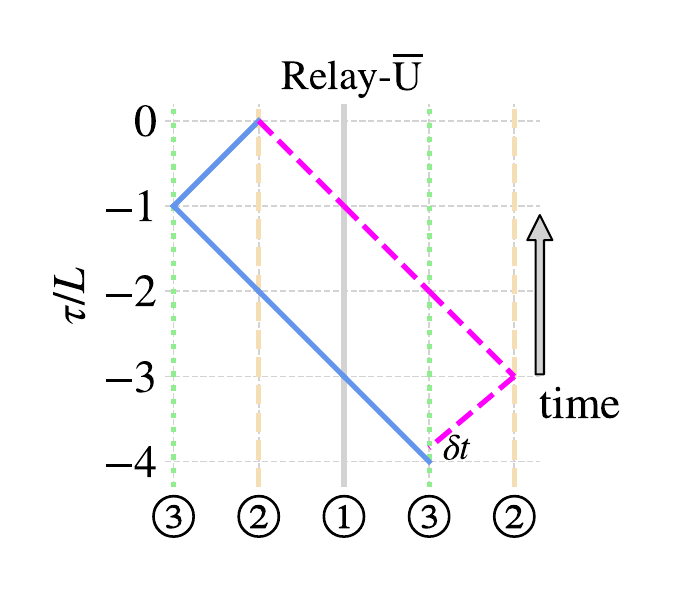} 
\caption{\label{fig:1st_tdi_diagram} The geometric diagrams for selected first-generation TDI channels. The vertical lines indicate the trajectories of three S/C (\textcircled{$i$} ($i = 1, 2, 3$), and the horizon is the separation between S/C. The blue and magenta lines depict two groups of combined laser links TDI.  }
\end{figure}

The first-generation TDI configurations were developed to mitigate laser noise caused by static unequal arm interferometry, as specified in references \cite{1997SPIE.3116..105N,1999ApJ...527..814A,2000PhRvD..62d2002E}. These configurations consist of five groups: Sagnac ($\alpha,\  \beta, \ \gamma$), Michelson (X, Y, Z), Relay (U, V, W), Monitor (D, F, G), and Beacon (P, Q, R). Of particular relevance to this study are the Michelson and Relay, and their first observables are expressed as
\begin{equation} \label{eq:X_U_measurement}
\begin{aligned}
{\rm X}  =& \left(  \mathcal{D}_{31} \mathcal{D}_{13} \mathcal{D}_{21}  \eta_{12}   + \mathcal{D}_{31} \mathcal{D}_{13}  \eta_{21}  + \mathcal{D}_{31} \eta_{13} + \eta_{31} \right) \\
 & -  ( \eta_{21} + \mathcal{D}_{21} \eta_{12} + \mathcal{D}_{21} \mathcal{D}_{12} \eta_{31} + \mathcal{D}_{21} \mathcal{D}_{12} \mathcal{D}_{31} \eta_{13} ), \\
\mathrm{U} = & ( \eta_{13} + \mathcal{D}_{13} \eta_{21} + \mathcal{D}_{21}  \mathcal{D}_{13}  \eta_{32} + \mathcal{D}_{32}  \mathcal{D}_{21}  \mathcal{D}_{13}  \eta_{23} )  \\
         & - (\eta_{23} + \mathcal{D}_{23} \eta_{32} + \mathcal{D}_{32}  \mathcal{D}_{23}  \eta_{13} + \mathcal{D}_{13} \mathcal{D}_{23}  \mathcal{D}_{32}  \eta_{21} ),
\end{aligned}
\end{equation}
where $\mathcal{D}_{ij}$ is a delay operator, $\mathcal{D}_{ij} y = y(t - L_{ij})$, $\eta_{ij}$ is measurement combination in a laser link from S/C$i$ to S/C$j$ as specified in \cite{Wang:2024a}. Three additional observables,  ($\mathrm{\bar{U}, \ \bar{V}, \ \bar{W}}$), could be derived by adjusting the path of Relay, and the first channel could be formulated as
\begin{equation} \label{eq:Ubar_measurement}
\begin{aligned}
\mathrm{\bar{U}} (t) =& (\eta_{32} + \mathcal{D}_{32} \eta_{23} + \mathcal{D}_{32}  \mathcal{D}_{23}  \eta_{12} + \mathcal{D}_{32} \mathcal{D}_{23}  \mathcal{D}_{12}  \eta_{31} )  \\
         & - ( \eta_{12} + \mathcal{D}_{12} \eta_{31} + \mathcal{D}_{12}  \mathcal{D}_{31}  \eta_{23} + \mathcal{D}_{12}  \mathcal{D}_{31}  \mathcal{D}_{23}  \eta_{32}  ).   
\end{aligned}
\end{equation}
More lucid expressions of three TDI observables are illustrated through geometric diagrams in Fig. \ref{fig:1st_tdi_diagram} \cite{Wang:2011,Wang:2020pkk}. The vertical lines represent the trajectories of three S/C (\textcircled{$i$} ($i = 1, 2, 3$), and duplicated S/C2 and S/C3 are plotted in the Relay-U and -$\mathrm{\bar{U}}$ diagrams to prevent intersections around non-integer delay ticks. The blue and magenta lines depict the combined links for the virtual interferometer of two beams. The $\delta t$ represents the mismatches for the two beams due to the orbital dynamics. 

\begin{figure}[htb]
\includegraphics[width=0.20\textwidth]{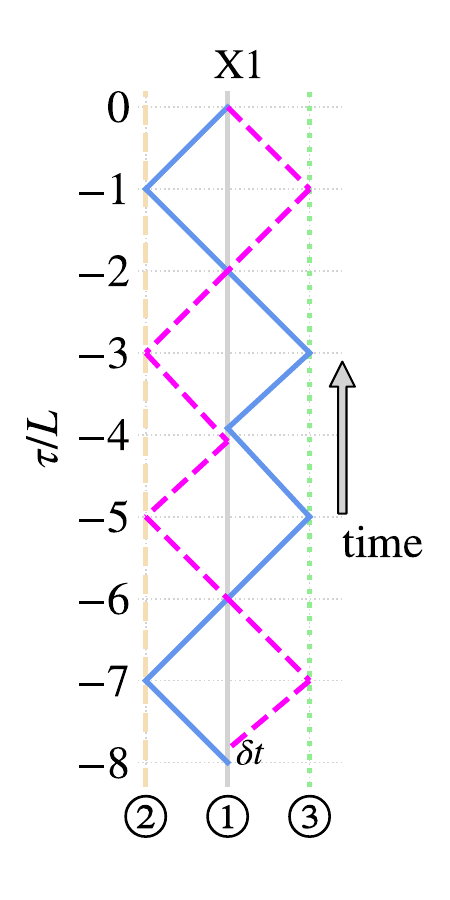}
\includegraphics[width=0.27\textwidth]{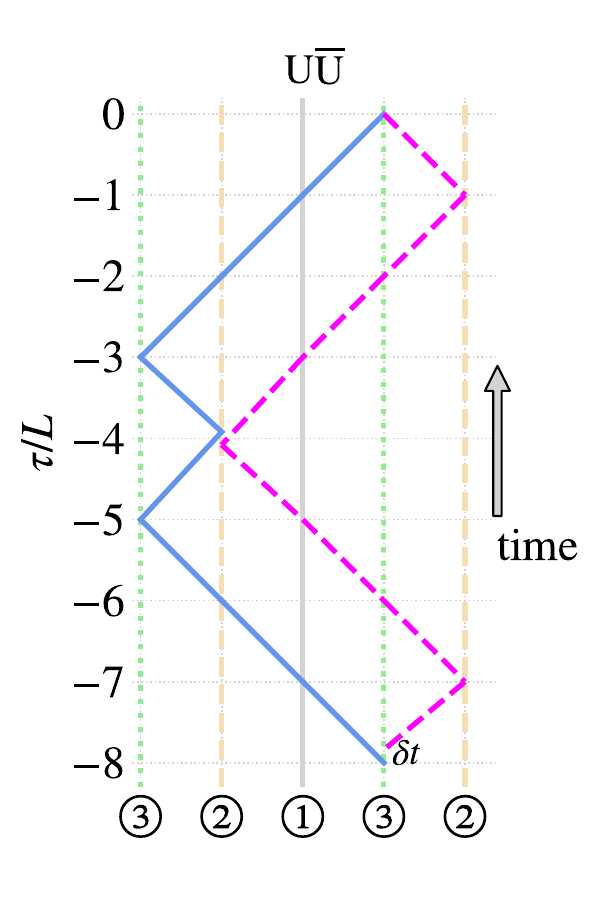} 
\caption{\label{fig:2nd_tdi_diagram} The geometric diagrams for X1 and $\mathrm{U\bar{U}}$ \cite{Wang:2011,Wang:2020pkk}.}
\end{figure}

Numerous second-generation TDI observables can be constructed from first-generation or other methods \cite{Vallisneri:2005ji,Hartwig:2021mzw,Tinto:2022zmf}. As the fiducial configuration, the second-generation TDI Michelson observables (X1, Y1, Z1) are derived by combining two of its first-generation TDI observables with a proper delay. For instance, X1 could be expressed as
\begin{equation} \label{eq:X1_TDI}
\mathrm{X1} (t) \simeq \mathrm{X}( t - 4 L) -  \mathrm{X}( t ),
\end{equation}
and its diagram is shown in the left plot of Fig. \ref{fig:2nd_tdi_diagram}. The Y1 and Z1 could be obtained through cyclical permutation of links' indexes. The first hybrid Relay observable could be generated by combining U and delayed $\mathrm{\bar{U}}$, denoted as U$\mathrm{\bar{U}}$ \cite{Wang:2011,Wang:2020pkk}, and vice versa. They can be expressed as
\begin{eqnarray}
\mathrm{U\bar{U}} (t) & \simeq  \mathrm{U}( t ) + \mathrm{\bar{U}}( t - 4 L ), \\
\mathrm{\bar{U}U} (t) & \simeq  \mathrm{\bar{U}}( t ) + \mathrm{U}( t - 4 L ).
\end{eqnarray}
The diagram of U$\mathrm{\bar{U}}$ is shown in the right plot of Fig. \ref{fig:2nd_tdi_diagram}.
These observables could further suppress the laser noise during S/C motion and should be categorized as a second-generation TDI.
Triple observables originating from three different S/C will be denoted as (U$\mathrm{\bar{U}}$, V$\mathrm{\bar{V}}$, W$\mathrm{\bar{W}}$) or ($\mathrm{\bar{U}}$U, $\mathrm{\bar{V}}$V, $\mathrm{\bar{W}}$W). Considering the performances of these two sets are effectively identical, the former group is elected for comparison with the fiducial Michelson observables (X1, Y1, Z1). 

The hybrid Relay is chosen to minimize null frequencies as a second-generation TDI configuration and is applied to reduce clock jitter with current method. 
Second-generation TDI channels are typically formed by combining two identical first-generation TDI channels with a time shift \cite{Shaddock:2003dj,Tinto:2022zmf}. These configurations introduce null frequencies at $n/(mL)$, where $n=1, 2, 3...$, $L$ is the arm length, and $mL$ is the relative time shift between two combined TDI paths. For instance, the time shift in X1 is $4L$ as exemplified in Eq. \eqref{eq:X1_TDI}, resulting in null frequencies at $n/(4L)$, as shown in next section. More second-generation TDI observables could be composed by synthesizing first-generation TDI channels at different S/Cs as developed in \cite{Vallisneri:2005ji}, and the interferometric beams are assembled with multiple beams. These observables may require a modified method for clock-jitter noise reduction, as current clock noise mitigation is applicable for two continuous beams \cite{Hartwig:2020tdu}. The hybrid Relay is combined with two distinct Relay channels, which can counteract their path mismatches, thus avoiding additional null frequencies due to time shift. Additionally, it can reduce clock noise using the current algorithm without any modification, as demonstrated in following section.

From three ordinary TDI observables $(a, b, c)$, three (quasi-)orthogonal or optimal observables, (A, E, T) can be constructed \cite{Prince:2002hp},
\begin{equation} \label{eq:abc2AET}
\begin{bmatrix}
\mathrm{A}  \\ \mathrm{E}  \\ \mathrm{T}
\end{bmatrix}
 =
\begin{bmatrix}
-\frac{1}{\sqrt{2}} & 0 & \frac{1}{\sqrt{2}} \\
\frac{1}{\sqrt{6}} & -\frac{2}{\sqrt{6}} & \frac{1}{\sqrt{6}} \\
\frac{1}{\sqrt{3}} & \frac{1}{\sqrt{3}} & \frac{1}{\sqrt{3}}
\end{bmatrix}
\begin{bmatrix}
a \\ b  \\ c
\end{bmatrix}.
\end{equation}
The A and E are expected to the science channels which can respond to GW signals effectively, while the T channel will be a noise-dominated data stream used to characterize instrumental noises.
To distinguish the optimal channels from different TDI configurations, a subscript of the first ordinary TDI is added to A/E/T. For instance, $\mathrm{A_{X1}}$ indicates the A channel generated from (X1, Y1, Z1).

\section{Noise suppressions and sensitivity} \label{sec:sensitivity}

\subsection{Suppressions of laser noise and clock noise} \label{subsec:laser_clock_noises}

\begin{figure}[thb]
\includegraphics[width=0.48\textwidth]{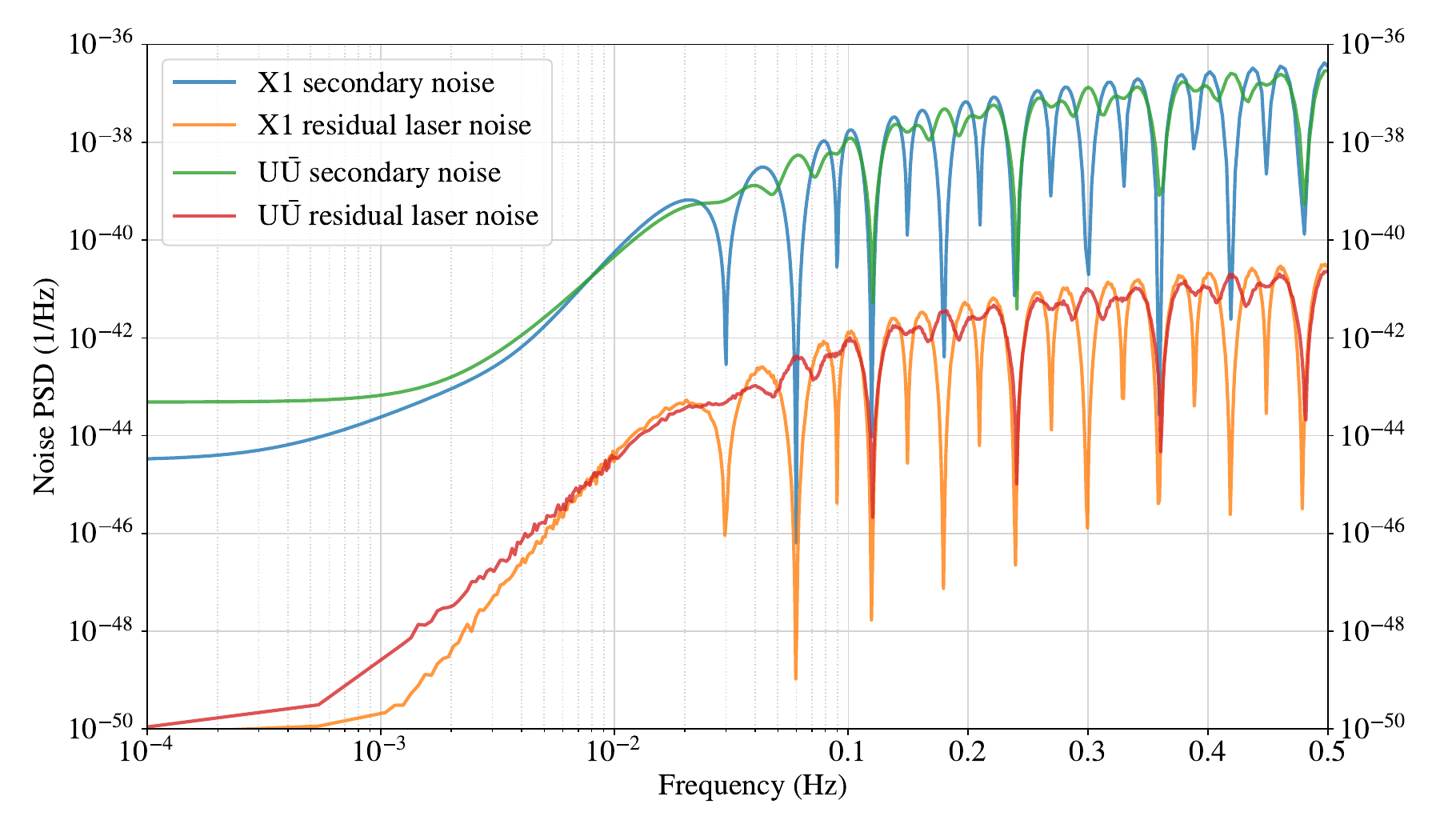}
\includegraphics[width=0.48\textwidth]{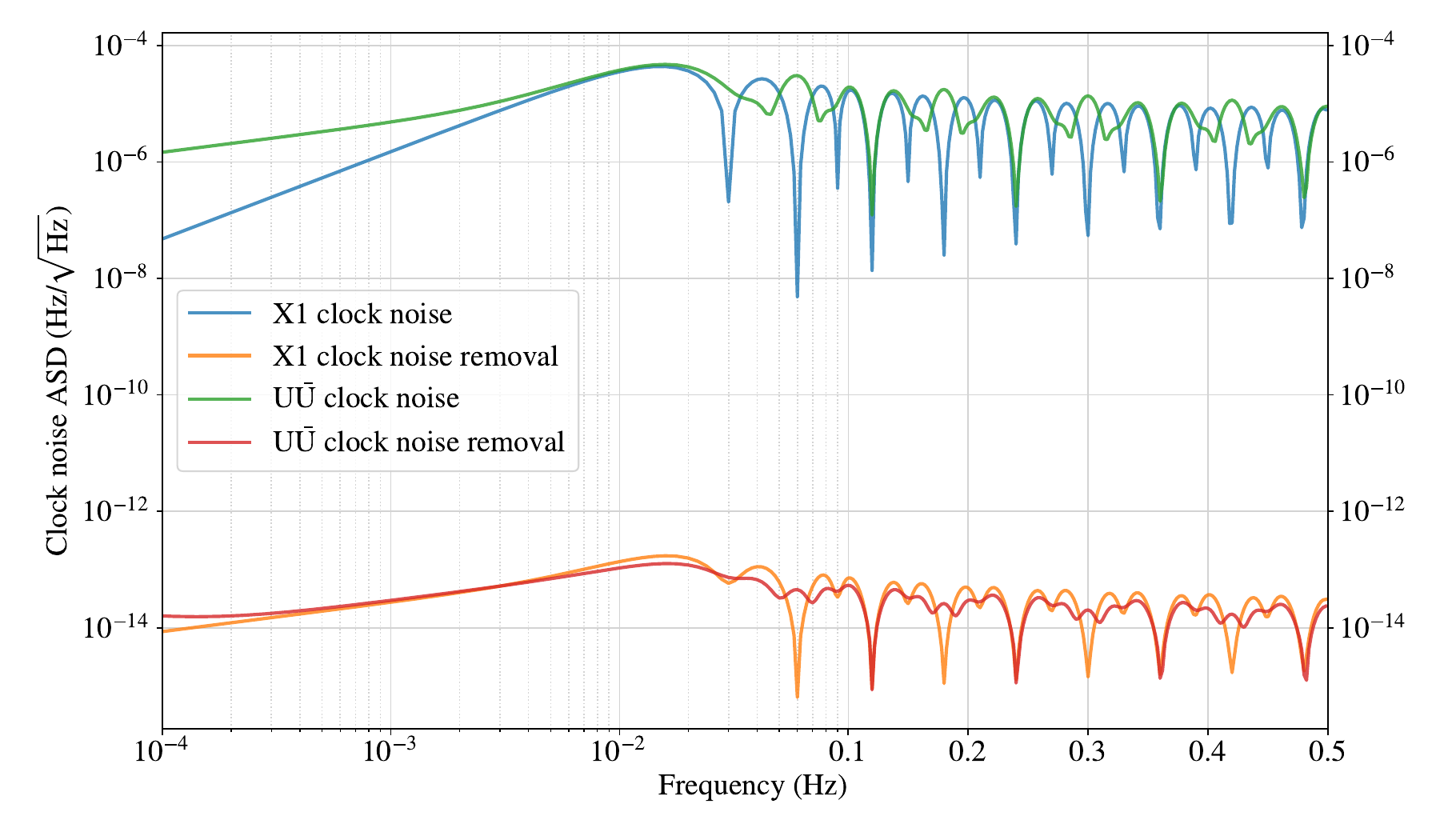} 
\caption{\label{fig:laser_clock_noise} The laser noise (upper) and clock noise (lower) suppressions in X1 and $\mathrm{U\bar{U}}$ observables. (A log scale is used for frequencies lower than 0.1 Hz, and a linear scale is employed for frequencies higher than 0.1 Hz. This mixed x-axis scale is utilized in the following frequency-domain plots). }
\end{figure}

The primary objective of TDI is to suppress the dominant laser noises. To assess the effectiveness of TDI observables, the lasers are assumed to have a stability of $30 \ \mathrm{Hz}/\sqrt{\mathrm{Hz}}$ which corresponds to a noise power-spectral density (PSD) of $1 \times 10^{-26}$/Hz. 
By employing a numerical orbit for LISA, the mismatches, $\delta t$, for X1 and U$\mathrm{\bar{U}}$ are within the ranges of [-10, 10] ps and [-6, 6] ps, respectively  \cite{Wang:2017aqq,Wang:2020a,Wang:2020pkk}. These mismatches are sufficiently small to theoretically meet the requirement for laser noise suppression. However, another critical factor influencing laser noise cancellation is the ranging error between two S/C. The post-processing ranging time between S/C could have a noise with PSD of $10^{-15} \frac{\mathrm{Hz}}{f} \ \mathrm{s/\sqrt{Hz}}$ and an additional bias of 3 ns (0.9 m) \cite{Bayle:2022okx,Staab:2023qrb}. With these noise assumptions, the residual laser noises in X1 and U$\mathrm{\bar{U}}$ are evaluated and shown in the upper panel of Fig. \ref{fig:laser_clock_noise}. The secondary noises, including acceleration noise and optical metrology noise, are plotted in the plot for comparison.
The noise budgets for acceleration noise $\mathrm{ S_{acc} }$ and optical path noise $\mathrm{S_{op}}$ are as follows \cite{2017arXiv170200786A},
\begin{equation}
\begin{aligned}
& \sqrt{ \mathrm{ S_{acc} } } = 3 \frac{\rm fm/s^2}{\sqrt{\rm Hz}} \sqrt{1 + \left(\frac{0.4 {\rm mHz}}{f} \right)^2 }  \sqrt{1 + \left(\frac{f}{8 {\rm mHz}} \right)^4 }, \\
& \sqrt{ \mathrm{ S_{op} } } = 10 \frac{\rm pm}{\sqrt{\rm Hz}} \sqrt{1 + \left(\frac{2 {\rm mHz}}{f} \right)^4 }.
 \end{aligned}
\end{equation}
From the plot, it is evident that the residual laser noises in two TDI channels are orders of magnitude lower than the secondary noises, indicating that the TDI solutions are effective for laser noise suppressions.

After the laser noise is suppressed, the clock noise becomes the key noise affecting sensitivity. Following the method developed in \cite{Hartwig:2020tdu}, clock noise is assessed for both X1 and U$\mathrm{\bar{U}}$ by assuming a clock noise of $4 \times 10^{-27} / f $ in fractional frequency deviations.
The results, depicted in the lower panel of Fig. \ref{fig:laser_clock_noise}. demonstrate that the clock noise can be significantly reduced for both Michelson and hybrid Relay. Hence, the current approach to clock noise cancellation is applicable to the new TDI configuration. 

From Fig. \ref{fig:laser_clock_noise}, we can notice that CFs of X1 for either laser noise, secondary noise, or (original) clock noise emerge at $n/(4L)\simeq 0.03 n \ \mathrm{Hz},\ n=1, 2, 3,...$ and $L$ is the arm length of constellation. In contrast, the CFs of U$\mathrm{\bar{U}}$ only appear at the $n/L \simeq 0.12 n$ Hz. With only one-fourth of CFs, the noise spectra of hybrid Relay are smoother and more stable than those of Michelson On the other side, the frequencies $f_c=n/L$ imply that the noise/signal cancellation occurs with a factor of $\sin \pi f L$ resulting from by a relative time delay of multiple $L$. Except the ordinary channels of first-generation TDI Sagnac, this factor commonly included in the noise PSD and GW response function of TDI channels, and we deduce that $f_c=n/L$ would be the minimum CFs for these TDI observables. And hybrid Relay observables (U$\mathrm{\bar{U}}$, V$\mathrm{\bar{V}}$, W$\mathrm{\bar{W}}$) only have these minimum CFs.  

\subsection{Average sensitivities}

\begin{figure}[htb]
\includegraphics[width=0.48\textwidth]{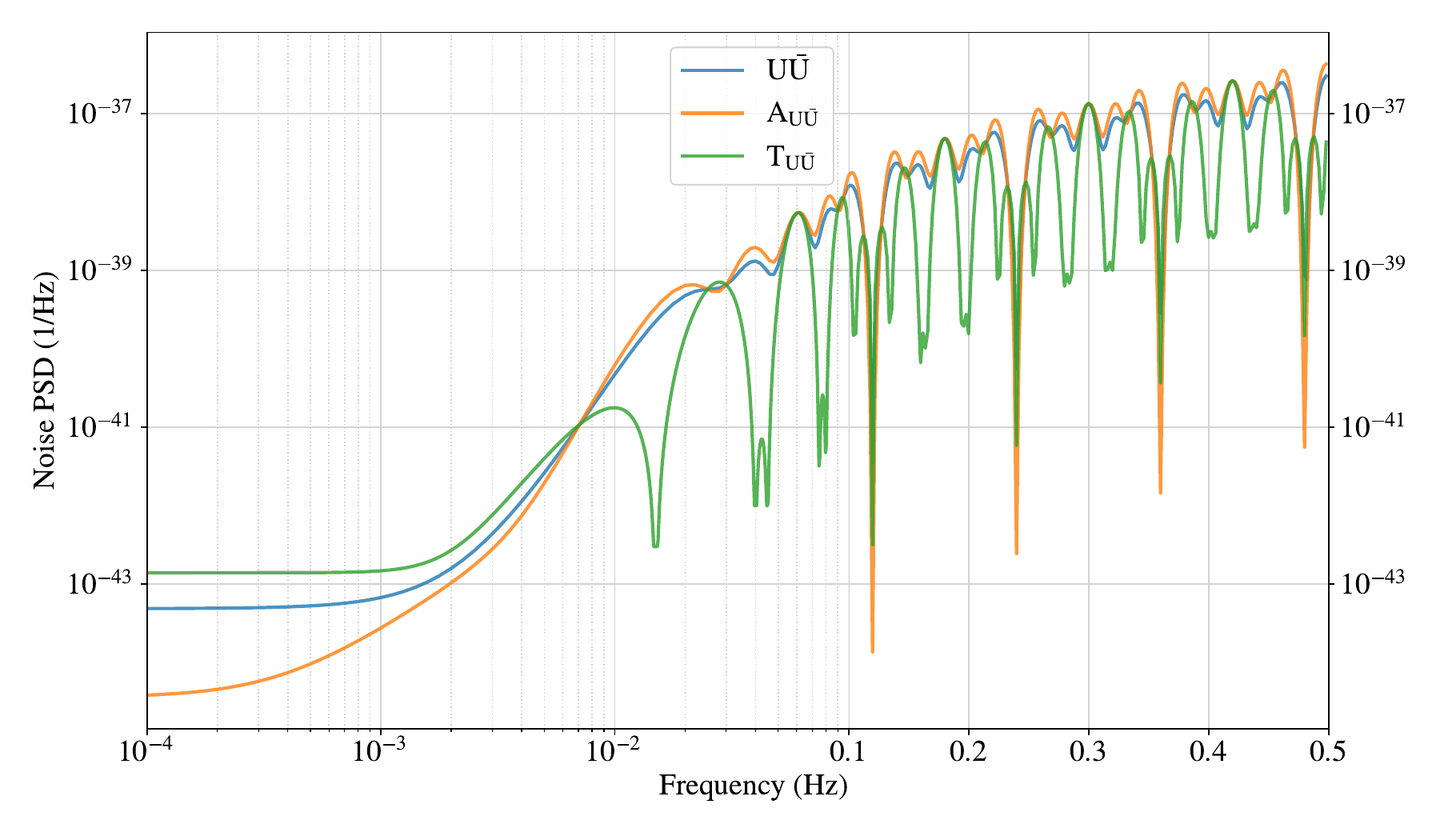}
\includegraphics[width=0.48\textwidth]{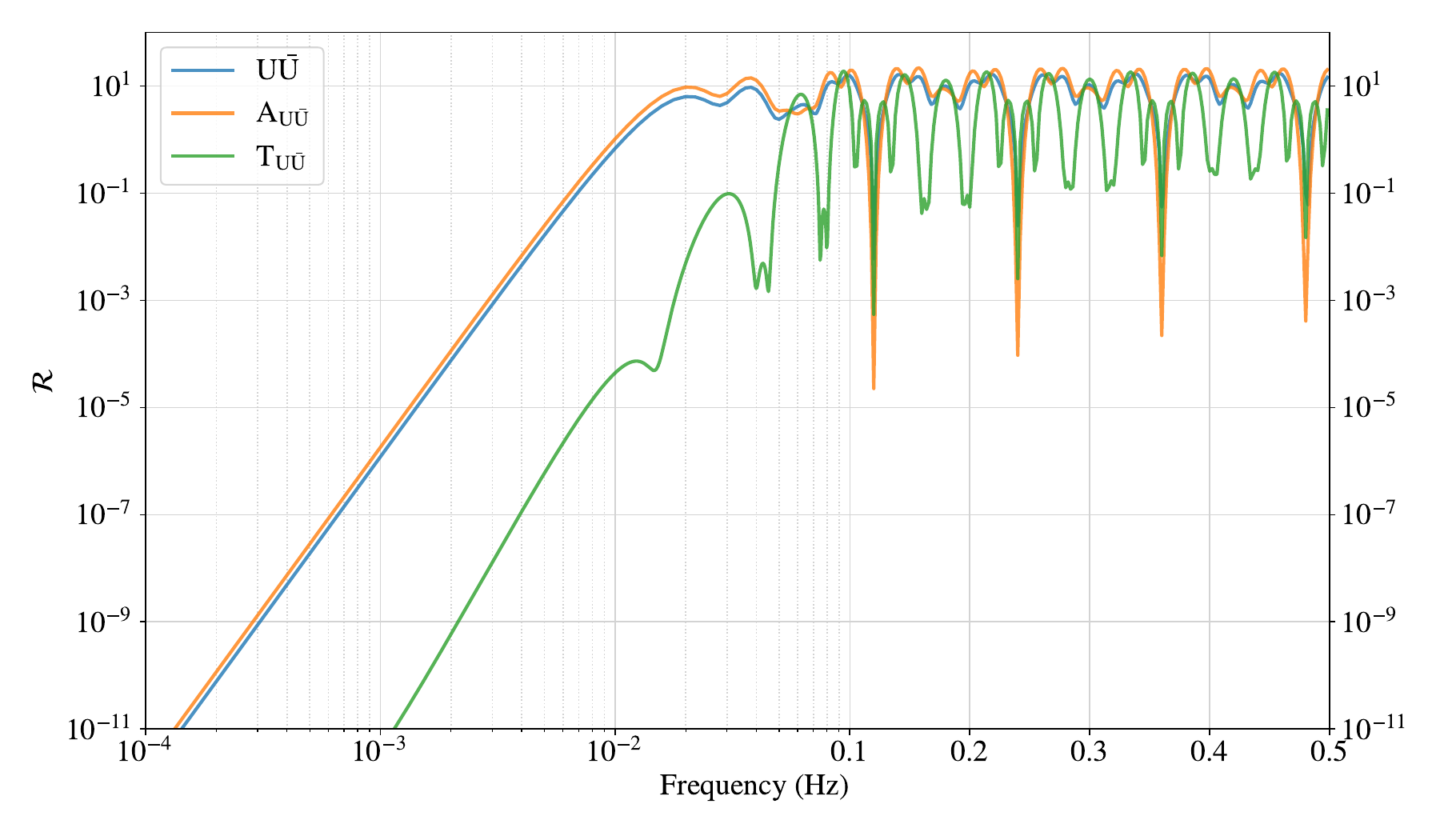}
\includegraphics[width=0.48\textwidth]{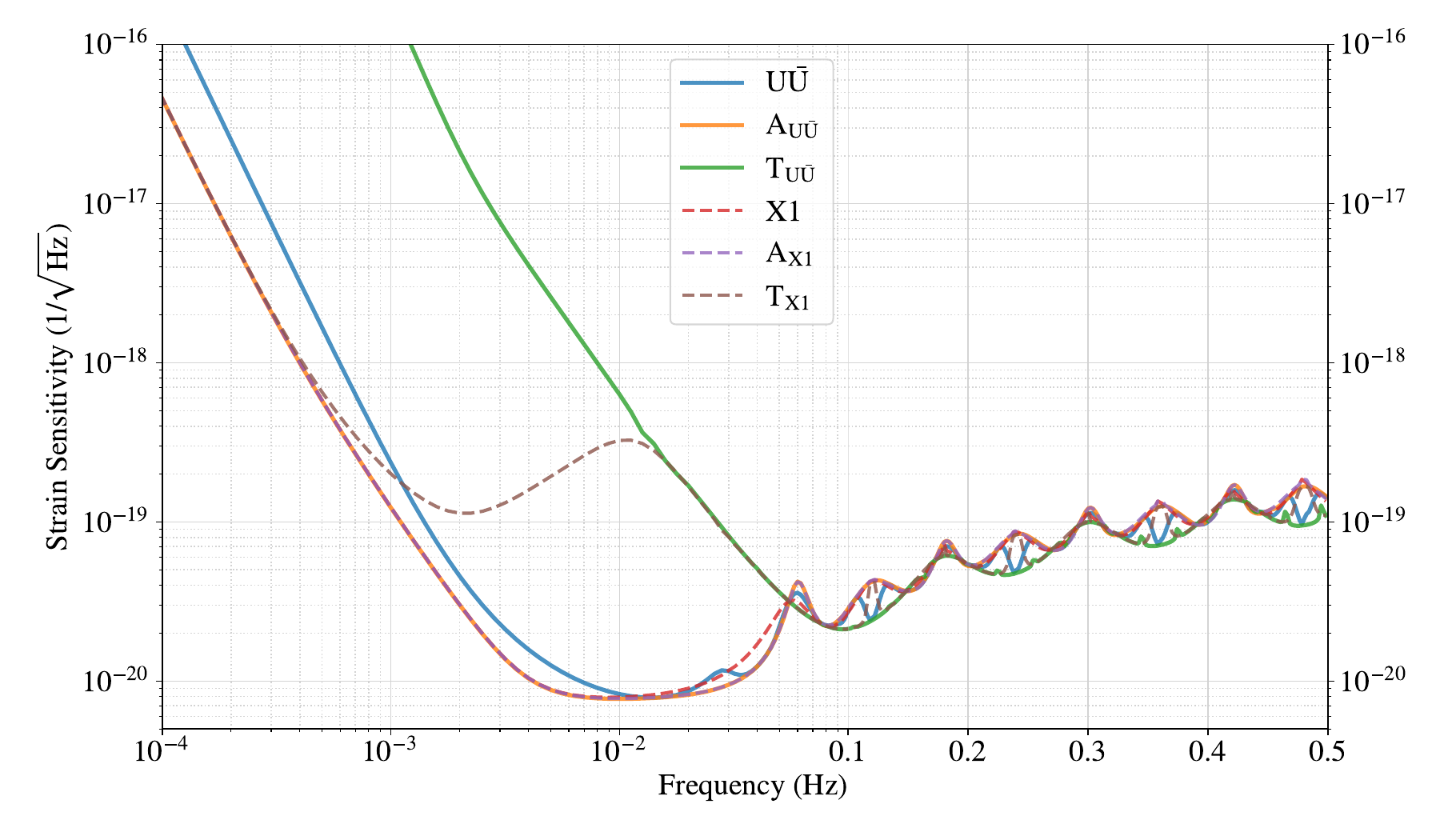}
\caption{\label{fig:sensitivity_UUbar} The noise PSDs (upper), average GW responses over sky directions (middle), and average sensitivities (lower) of hybrid Relay TDI channels. An average sensitivity is obtained by weighting the noise PSD with the corresponding GW response, $\sqrt{\mathrm{PSD}/\mathcal{R}}$.}
\end{figure}

The average sensitivity over all sky directions indicates the detectability of a TDI observable. The sensitivities of U$\mathrm{\bar{U}}$ and the corresponding optimal channels are evaluated by using the algorithm in \cite{Wang:2020pkk}, and the results are shown in Fig. \ref{fig:sensitivity_UUbar}. The PSDs of secondary noises are depicted in the upper panel, and the coefficients of acceleration noises and optical metrology noises for the PSD and cross-spectral density (CSD) are provided in Table \ref{tab:TDI_PSD} and \ref{tab:TDI_CSD} in the Appendix. The separated illustrations of optical metrology noise and acceleration noise in three TDI channels are presented in Fig. \ref{fig:psd_UUbar_inst}. As aforementioned, the CFs of U$\mathrm{\bar{U}}$ and A$_\mathrm{U\bar{U}}$/E$_\mathrm{U\bar{U}}$ occur around integral multiples of $1/L=0.12$ Hz. In addition to these shared CFs, the T$_\mathrm{U\bar{U}}$ has extra CFs, with the lowest frequency being $1/(8L)\simeq 0.015$ Hz.

The averaged GW responses over all sky directions at a given time are depicted in the middle panel of Fig. \ref{fig:sensitivity_UUbar}. As observed from the plot, the CFs in the responses align with the noise PSD. In comparison to the other two observables, the response of T$_\mathrm{U\bar{U}}$ is significantly poorer for frequencies lower than 50 mHz, and the response becomes comparable at higher frequencies. The average sensitivities are obtained by weighting the noise PSD with GW response, $\sqrt{\mathrm{PSD}/\mathcal{R}}$, and the curves are shown in the lower panel of Fig. \ref{fig:sensitivity_UUbar}. We emphasize that these sensitivities are instantaneous, the noise PSD and the response are calculated at a selected time point, the spikes of CFs in the PSDs and responses could counteracted numerically. In a realistic scenario, the noise PSD is estimated from a chunk of data during the detector's orbit, and the actual response to a specific source varies with time, resulting in time-dependent sensitivities.

The sensitivities of fiducial Michelson TDI channels are also calculated and depicted using dashed curves for comparison. The sensitivities of A$_\mathrm{U\bar{U}}$ and A$_\mathrm{X1}$ are identical, and the sensitivities of E channels from two TDI configurations are consistent with their A channels. However, the sensitivities of T differ from the science channels. In the frequency band higher than 50 mHz, the T$_\mathrm{U\bar{U}}$ could be slightly more sensitive than other observables. Due to the unequal arm, T$_\mathrm{X1}$ diverges from T$_\mathrm{U\bar{U}}$ at 15 mHz and becomes as sensitive as the A/E for frequencies lower than 0.3 mHz, as analyzed in \cite{Adams:2010vc,Wang:2020fwa}.

\section{Simulation and analysis} \label{sec:simulation_analysis}

\subsection{Data simulation}

The data simulation is performed by using \textsf{SATDI} \cite{Wang:2024a}. Noise data are generated with a sampling rate of 4 Hz for 30 days, and TDI is implemented by using interpolation \cite{Shaddock:2004ua}. The noise sources include acceleration noises and optical metrology noise by assuming that laser noise and clock noise are sufficiently suppressed. Simulations are carried out for second-generation Michelson (X1, Y1, Z1) and hybrid Relay (U$\mathrm{\bar{U}}$, V$\mathrm{\bar{V}}$, W$\mathrm{\bar{W}}$). Optimal observables are generated by using Eq. \eqref{eq:abc2AET} and employed to evaluate their analytical capabilities. 

The GW signals of MBBH are calculated for 25 days preceding their coalescences. Two sources are simulated with different masses: source1 is a binary with $m_1 = 6 \times 10^5 M_\odot$ and $m_2 = 10^5 M_\odot$ in the source frame, and source2 is less massive with $m_1 = 1.2 \times 10^4 M_\odot$ and $m_2 = 2 \times 10^3 M_\odot$. The GW signal of source1 evolves from 0.2 mHz to 20 mHz in 25 days in the detector frame, and the chirp of source2 lasts from 2.3 mHz to 0.5 Hz. These two sources correspond to the signals in the low- and high-frequency bands, respectively. The spins of the black holes are ignored in this simulation. Both two binaries are assumed to originate from an ecliptic longitude $\lambda=4.6032$ rad and ecliptic longitude $\beta=\pi/10$ rad at redshift $z=1$, corresponding to a luminosity distance of $d_L = 6791.8$ Mpc \cite{Planck:2018vyg,Astropy:2013muo}. The polarization angle is set to be $\psi=0.55$ rad, and the inclination angle between the binaries' rotation and the line-of-sight is $\iota=\pi/3$. The merger time $t_c$ is set to be at the 56th day of mission orbit with reference phase $\phi_c$. The time-domain waveforms are generated by using the SEOBNRv4HM model \cite{Bohe:2016gbl}. 

\begin{figure}[htb] 
\includegraphics[width=0.44\textwidth]{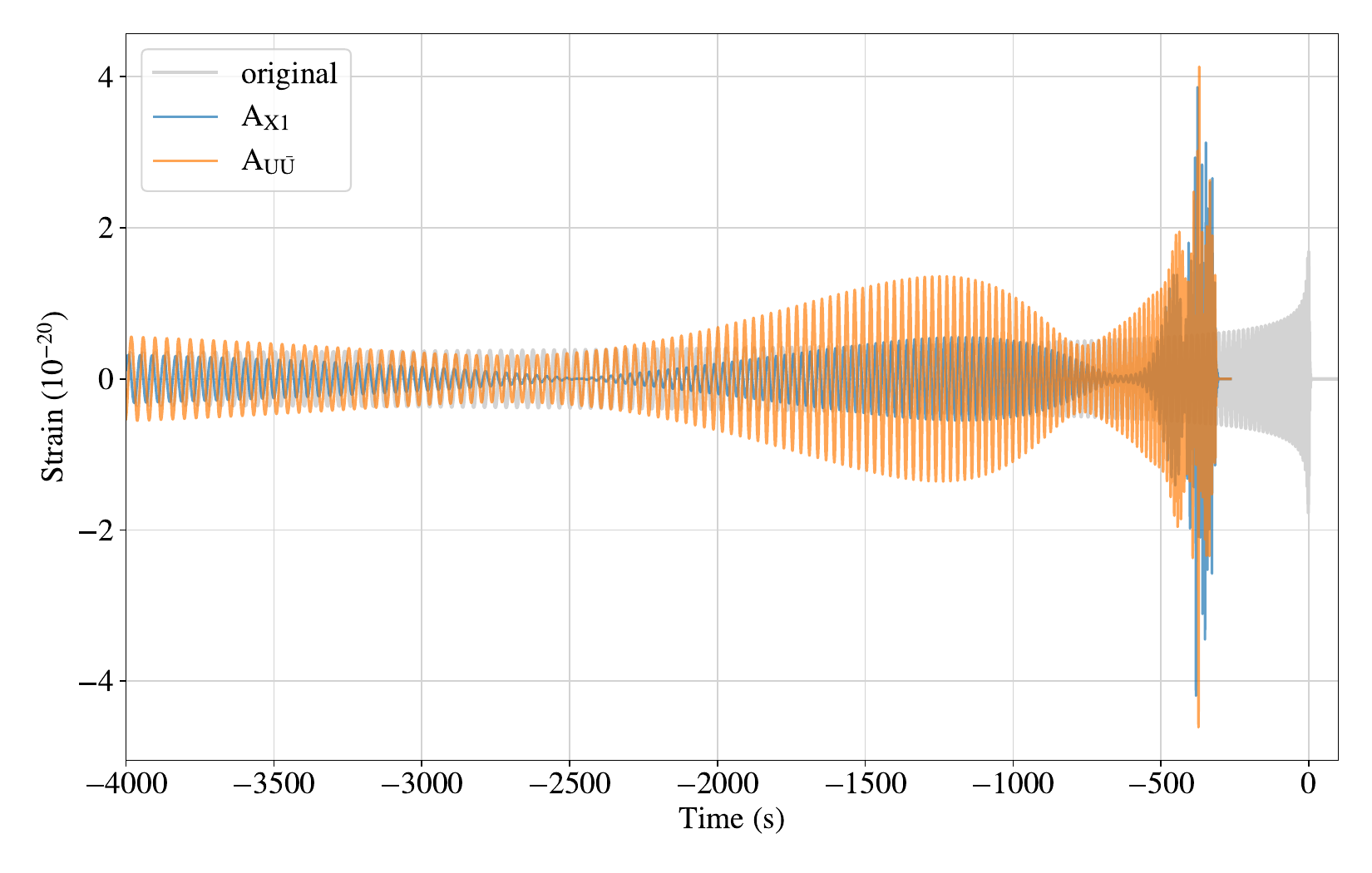}
\includegraphics[width=0.48\textwidth]{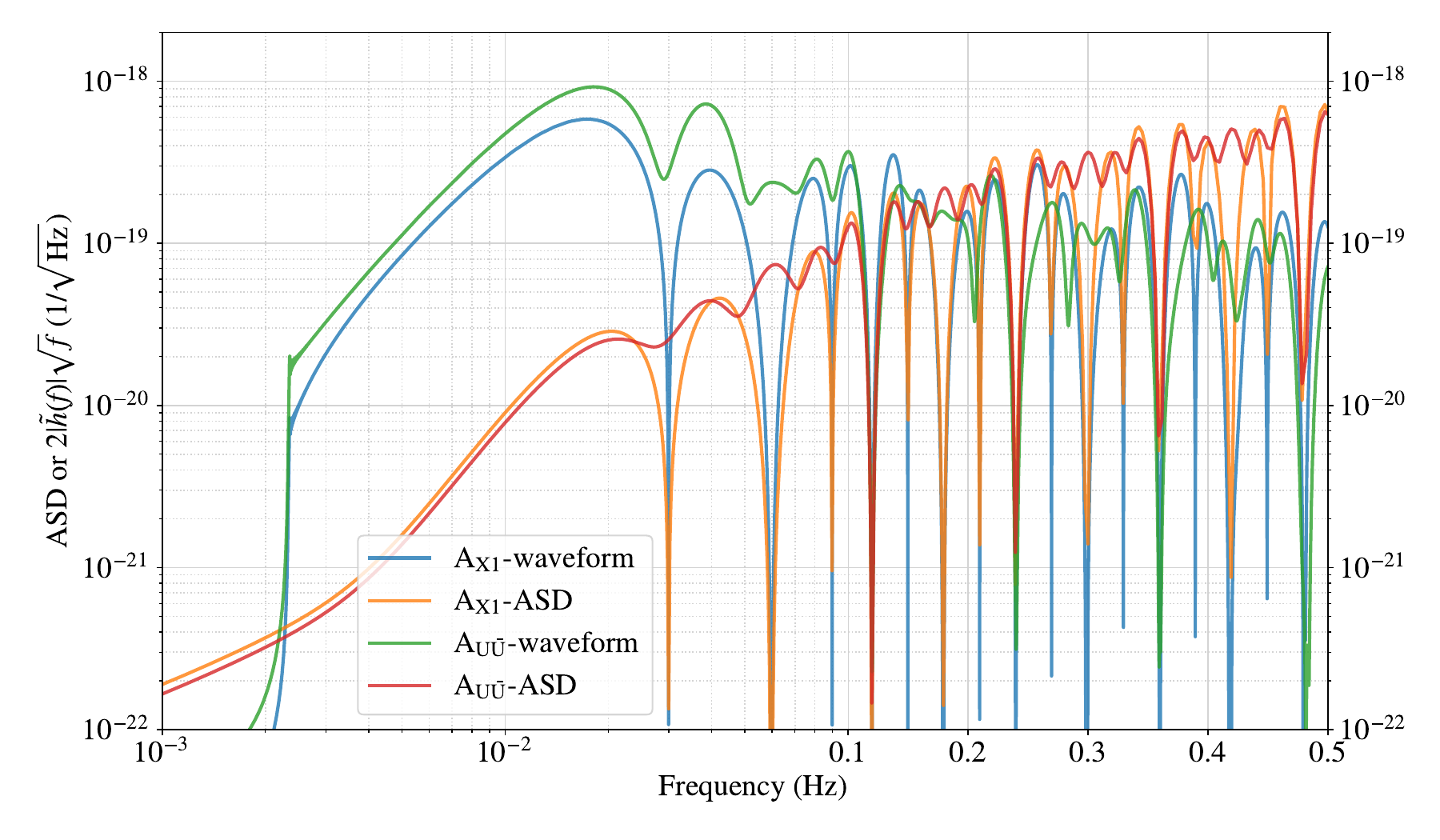}
\caption{\label{fig:waveform_td_fd} The time-domain (upper) and frequency-domain (lower) waveforms of source2 ($m_1 = 1.2 \times 10^4 M_\odot$, $m_2 = 2 \times 10^3 M_\odot$) in A$_\mathrm{X1}$ and A$_\mathrm{U\bar{U}}$. In the upper plot, the grey curve is the original GW signal arrived at the solar-system barycentric (SSB) center. The x-axis is the barycentric dynamical time with respect to the merger. In current simulation setups and source selection, the merger signal arrives at the detector earlier than it reaches SSB center.}
\end{figure}

The waveforms of source2 approaching the merger are depicted in Fig. \ref{fig:waveform_td_fd}. The x-axis is the barycentric dynamical time (TDB) with respect to the merger at the solar-system barycentric (SSB) center.
The grey curve illustrates the GW signal arriving at the SSB center, and the blue and orange curves represent the responded waveforms in A$_\mathrm{X1}$ and A$_\mathrm{U\bar{U}}$ observables, respectively. As displayed in the plot, the amplitudes of waveforms are modulated by the TDI response functions. In A$_\mathrm{X1}$, the signal amplitude is significantly suppressed around its CFs. Benefiting from the minimal CFs, the waveform in A$_\mathrm{U\bar{U}}$ is less restrained. Additionally, phase aliasing occurs at high frequencies due to the longer time spans of TDI compared to the GW periods.
Frequency-domain waveforms are obtained via Fourier transform and presented in the lower panel of Fig. \ref{fig:waveform_td_fd}. The noise PSDs of A$_\mathrm{X1}$ and A$_\mathrm{U\bar{U}}$ are plotted for comparison, revealing clearer suppression of waveforms at their CFs. In the Michelson observable A$_\mathrm{X1}$, the waveform amplitude quickly drops at each CF, as does its PSD. The waveform in A$_\mathrm{U\bar{U}}$ channel is subjected to a only few CFs.

\subsection{Parameter inferences}

The parameters of injected binaries are estimated by using Bayesian algorithm. The inferences are conducted utilizing the optimal observables from second-generation hybrid Relay (A$_\mathrm{U\bar{U}}$, E$_\mathrm{U\bar{U}}$, T$_\mathrm{U\bar{U}}$) and Michelson (A$_\mathrm{X1}$, E$_\mathrm{X1}$, T$_\mathrm{X1}$) to compare their performances. 
The likelihood function of parameter estimation is given by \cite{Romano:2016dpx}
\begin{equation}
\begin{split}
\ln \mathcal{L} (d|\vec{\theta} ) = \sum_{f_i} & \left[  
-\frac{1}{2} (\tilde{\mathbf{d}} - \tilde{\mathbf{h}} ( \vec{\theta}) )^T \mathbf{C}^{-1}  (\tilde{\mathbf{d}} - \tilde{\mathbf{h}}(\vec{\theta}) )^\ast \right. \\
 &  \left.  - \ln \left( \det{ 2 \pi \mathbf{C} } \right)
   \right],
\end{split}
\end{equation}
where $\tilde{\mathbf{d}}$ is frequency-domain TDI data obtained via Fourier transform, $\tilde{\mathbf{h}} (\vec{\theta})$ denotes modeled GW signal described by the parameter vector $\vec{\theta} \in [\mathcal{M}_c, q, d_L, \iota, \lambda, \beta, \psi, t_c, \phi_c]$. The responded waveform synthesizes the TDI response function and frequency-domain waveform generated from a reduced-order model of SEOBNRv4HM \cite{Cotesta:2020qhw}. The $\mathbf{C}$ is the correlation matrix of noises from three optimal channels,
\begin{equation} \label{eq:cov_mat}
 \mathbf{C} = \frac{T_\mathrm{obs}}{4}
 \begin{bmatrix}
S_\mathrm{AA} & S_\mathrm{AE} & S_\mathrm{AT} \\
S_\mathrm{EA} & S_\mathrm{EE} & S_\mathrm{ET} \\
S_\mathrm{TA} & S_\mathrm{TE} & S_\mathrm{TT}
\end{bmatrix},
\end{equation}
where $T_\mathrm{obs}$ is the duration of data, and PSD $S_{xx}$ and CSD $S_{xy}$ are estimated by using the algorithm detailed in \cite{Allen:2005fk}. The priors for parameters are provided in Table \ref{tab:priors}. The redshifted chirp mass $\mathcal{M}_c = (m_1 m_2)^{3/5}/(m_1 + m_2)^{1/5}$ and mass ratio $q=m_2/m_1$ are inferred instead of inferring individual component masses. The parameter estimations are implemented employing the nested sampler \textsf{MultiNest} \cite{Feroz:2008xx,Buchner:2014nha}. 

\begin{table}[tbh]
\caption{\label{tab:priors}
Priors for parameter estimation.}
\begin{ruledtabular}
\begin{tabular}{cccc}
Variable & Prior &  range & unit  \\
\hline
$\mathcal{M}_c$ & uniform &  & $M_\odot$  \\
$q$ & uniform & [0.1, 0.2] &  \\
$d_L$ & Comoving & [5000, 8000] & Mpc  \\
$\iota$ & $\sin$ & [0, $\pi$] & rad  \\
$\lambda$ & uniform & [0, $2 \pi$] & rad  \\
$\beta$ & $\cos$ & [$ -\frac{\pi}{2}, \ \frac{\pi}{2}$] & rad  \\
$\psi$ & uniform & [0, $\pi$] & rad  \\
$t_c$ & uniform & [$t_c$-180, $t_c$+180] & second  \\
$\phi_c$ & uniform & [0, $2 \pi$] & rad  \\
\end{tabular}
\end{ruledtabular}
\end{table}

\begin{figure*}[htb]
\includegraphics[width=0.95\textwidth]{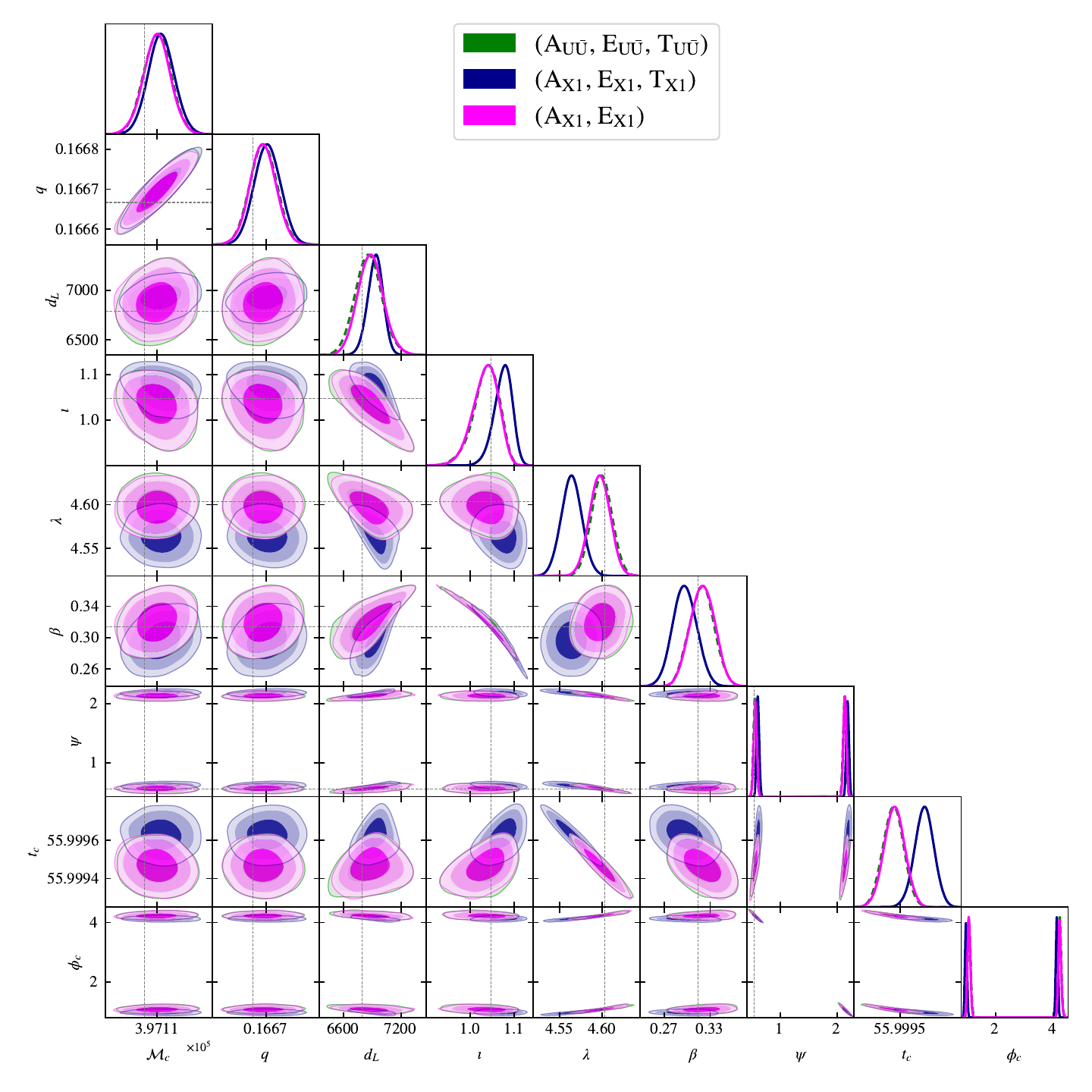}
\caption{\label{fig:corner_6p0e5} The inference results for source1 ($m_1 = 6 \times 10^5 M_\odot$ and $m_2 = 10^5 M_\odot$) from three TDI combinations. Three gradients show the $1\sigma$, $2\sigma$, and $3\sigma$ regions from darker to lighter. }
\end{figure*}

The inferred distributions for parameters of source1 are shown in Fig. \ref{fig:corner_6p0e5}, and results from three TDI combinations are presented. The green colors represent the inferred parameter distributions from (A$_\mathrm{U\bar{U}}$, E$_\mathrm{U\bar{U}}$, T$_\mathrm{U\bar{U}}$), the blue hues signify estimation from (A$_\mathrm{X1}$, E$_\mathrm{X1}$, T$_\mathrm{X1}$), and the magentas depict the result of (A$_\mathrm{X1}$, E$_\mathrm{X1}$). Among these three combinations, with the exception of (A$_\mathrm{X1}$, E$_\mathrm{X1}$, T$_\mathrm{X1}$), the inferences from other two combinations are consistent, and injected parameters fall within sensible ranges. The discrepancy between (A$_\mathrm{X1}$, E$_\mathrm{X1}$) and (A$_\mathrm{X1}$, E$_\mathrm{X1}$, T$_\mathrm{X1}$) suggests that the T$_\mathrm{X1}$ undermines the parameter estimation. 

\begin{figure*}[htb]
\includegraphics[width=0.95\textwidth]{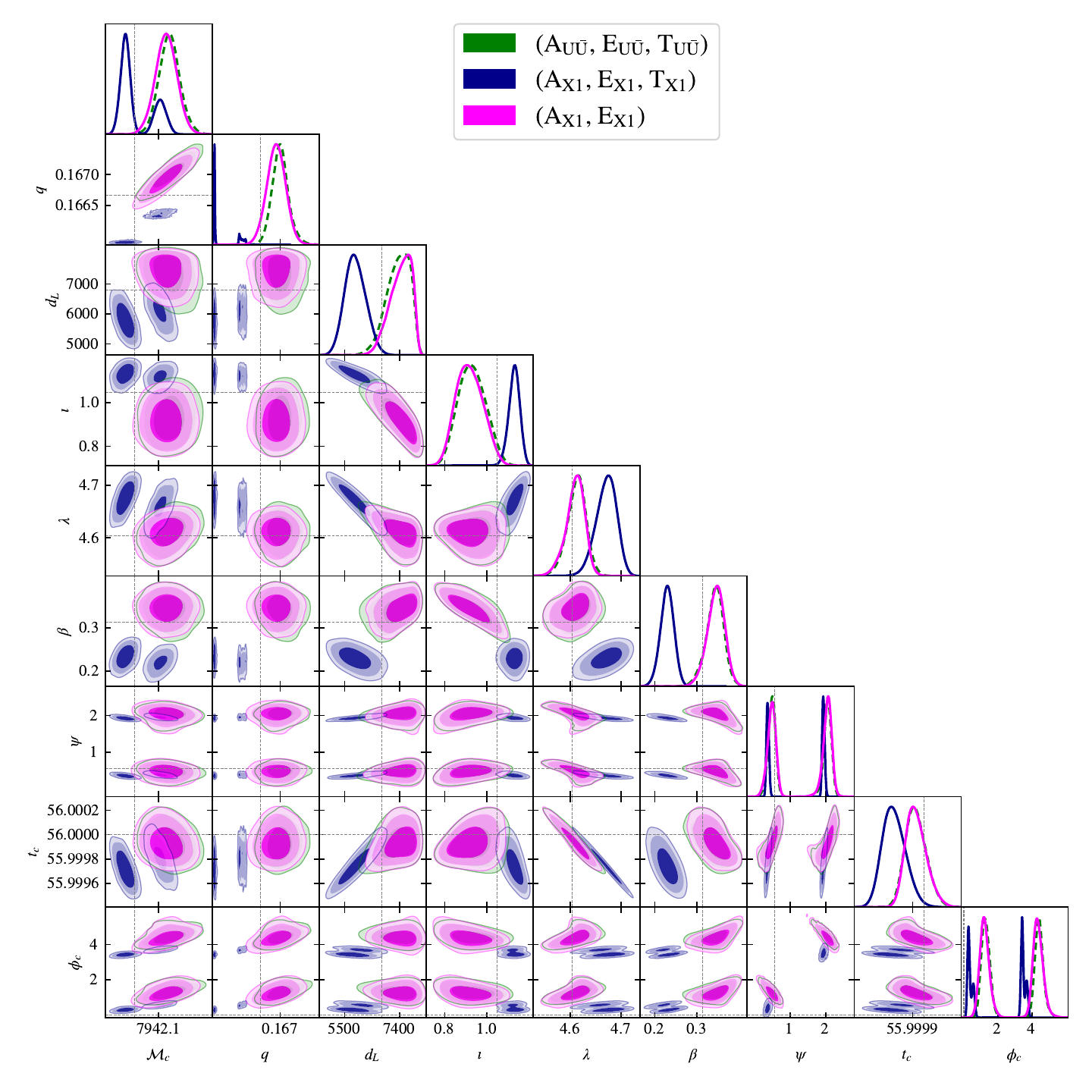}
\caption{\label{fig:corner_1p2e4} The corner plots for source2 ($m_1 = 1.2 \times 10^4 M_\odot$, $m_2 = 2 \times 10^3 M_\odot$) from three TDI combinations. Three gradients show the $1\sigma$, $2\sigma$ and $3\sigma$ regions from darker to lighter.}
\end{figure*}

The inferred distributions for source2 are illustrated in Fig. \ref{fig:corner_1p2e4}. The result from the dataset (A$_\mathrm{X1}$, E$_\mathrm{X1}$, T$_\mathrm{X1}$) noticeably deviates from the injected values. After the T$_\mathrm{X1}$ data is removed, the science data streams (A$_\mathrm{X1}$, E$_\mathrm{X1}$) yield a reasonable result that agrees with the distributions from (A$_\mathrm{U\bar{U}}$, E$_\mathrm{U\bar{U}}$, T$_\mathrm{U\bar{U}}$). This further confirms that T$_\mathrm{X1}$ data is problematic for signal analysis. Additionally, we can notice that inferred mass parameters ($\mathrm{M}_c,\ q$) are (almost) beyond the creditable regions for the (A$_\mathrm{U\bar{U}}$, E$_\mathrm{U\bar{U}}$, T$_\mathrm{U\bar{U}}$) group, and this could be caused by the discrepancies between the time-domain waveform model and frequency-domain waveform approximant \cite{Cutler:2007mi}. 
For more massive sources1 binary, its waveform has fewer cycles than source2 over the span of 25 days. As a result, the current faithfulness between time-domain waveform and frequency-domain waveform models may allow for reasonable estimation of the mass parameters for source1, but might not meet the matching requirements for estimating the masses of source2. In the results from two sources, the hybrid Relay demonstrates robust outcomes with its full optimal data set, while the Michelson configuration is susceptible to bias introduced by its T channel. When comparing the results of Michelson across two sources, it appears that source1 exhibits less bias than source2, possibly because the A and E exert greater contributions than the T channel in the low-frequency band.

\subsection{Post-analysis} \label{subsec:hindsight}

\begin{figure*}[htb]
\includegraphics[width=0.48\textwidth]{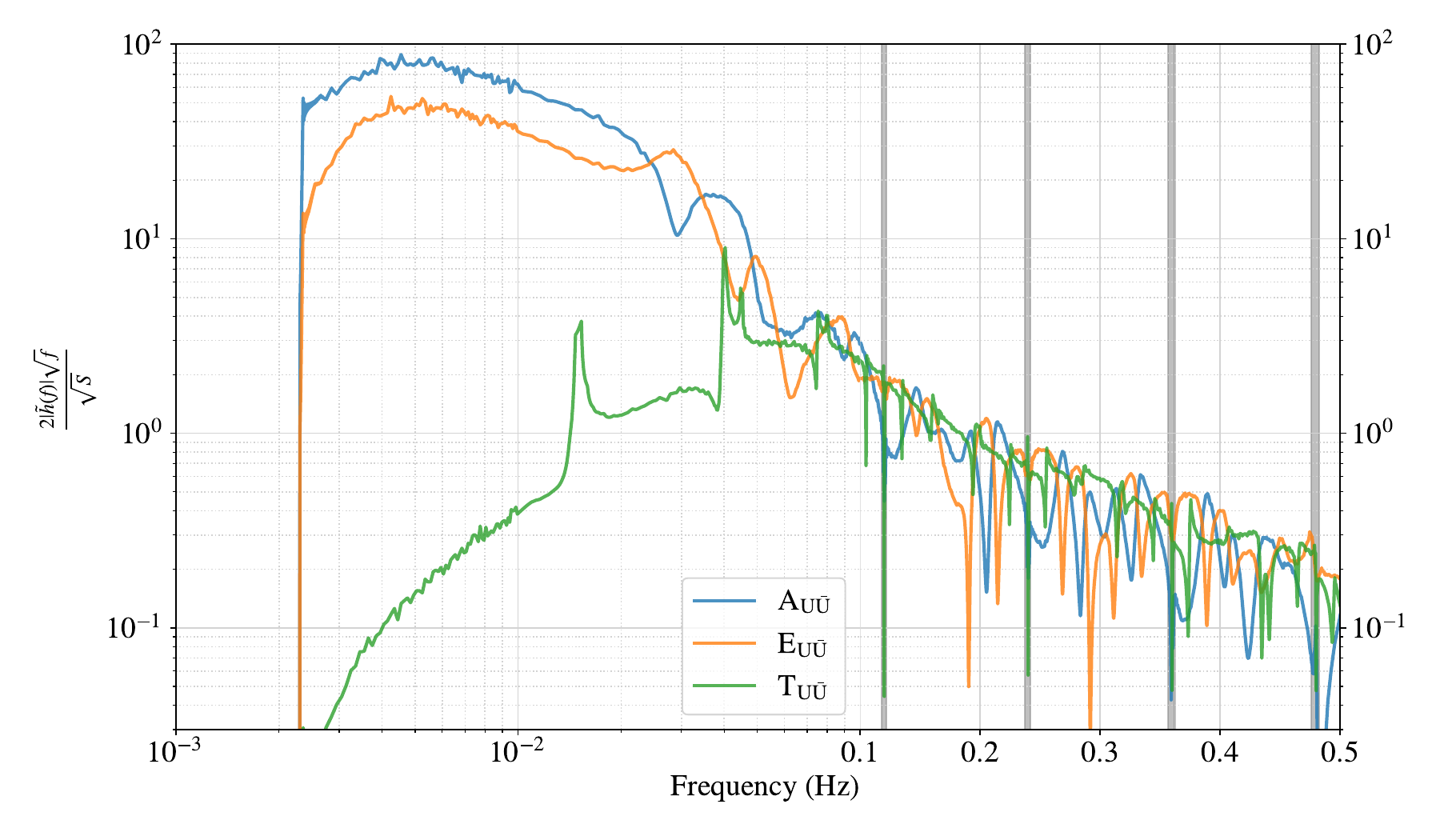}
\includegraphics[width=0.48\textwidth]{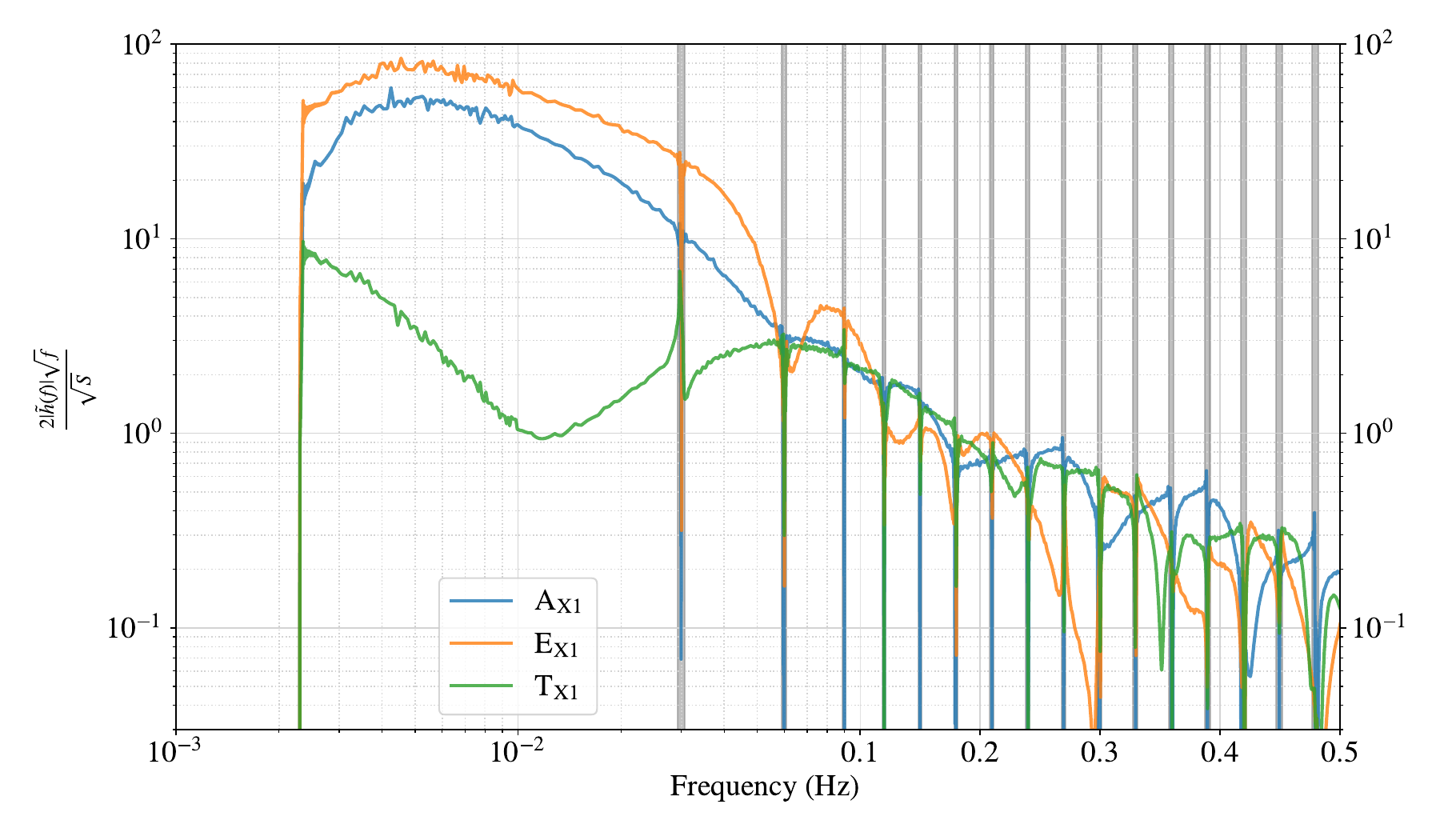}
\includegraphics[width=0.48\textwidth]{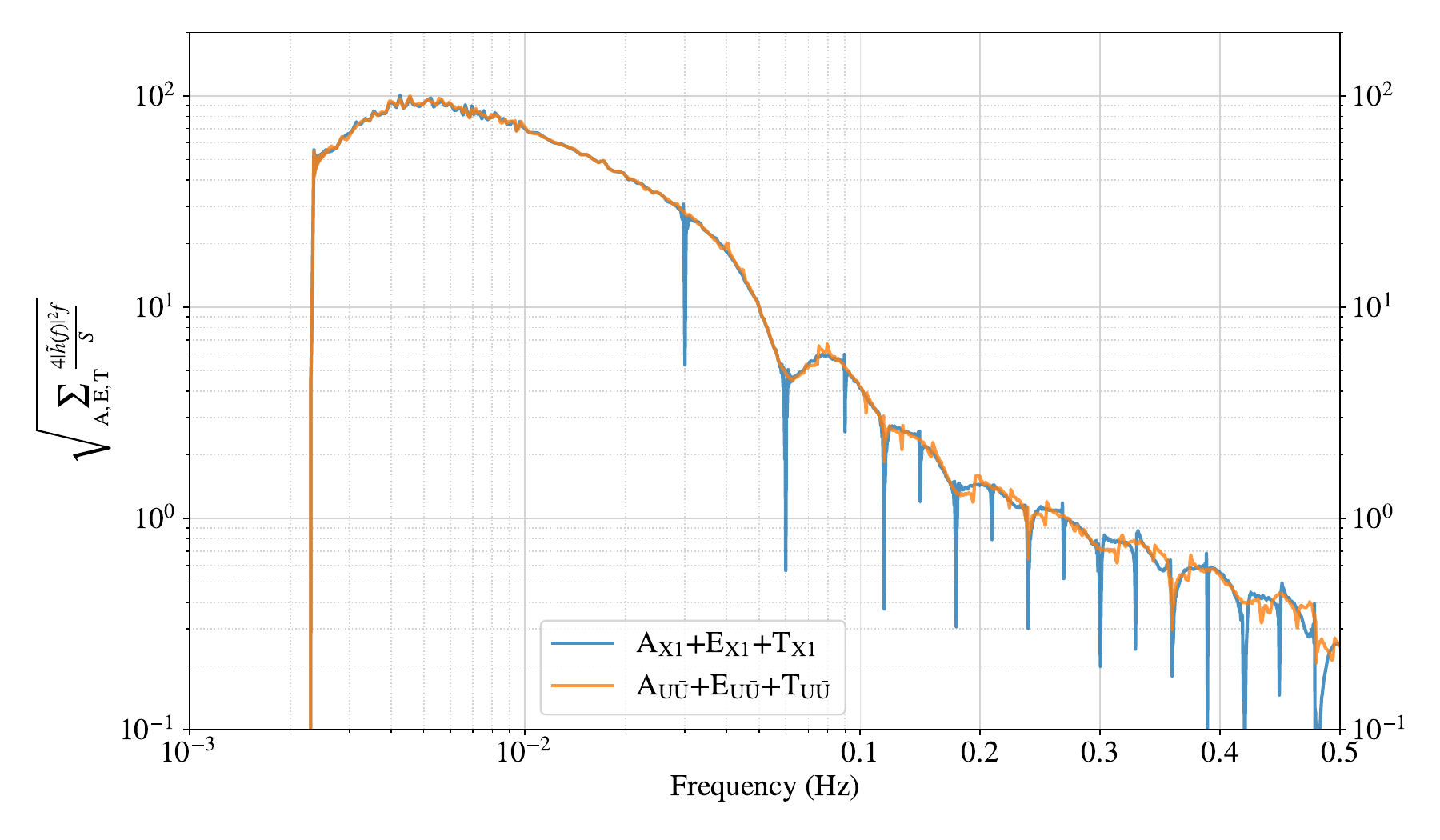}
\caption{\label{fig:delta_snr} The ratio $\frac{2 |\tilde{h}(f)| \sqrt{f}}{\sqrt{S}}$ for source2 in the optimal TDI channels. The upper left shows the ratios from (A$_\mathrm{U\bar{U}}$, E$_\mathrm{U\bar{U}}$, T$_\mathrm{U\bar{U}}$), the upper right is the result from (A$_\mathrm{X1}$, E$_\mathrm{X1}$, T$_\mathrm{X1}$), and their CFs are marked by grey vertical lines. The lower plot is the square root of quadratic sums of optimal data streams. (In principle, the plots should use a log-scale x-axis to reflect the SNR contributions at different frequencies, the linear scale is utilized at high frequencies to show the CFs more clearly.) }
\end{figure*}

A further investigation and comparison are performed for the Michelson and hybrid Relay TDI configurations. Three optimal channels utilized are (quasi-)orthogonal. By ignoring their cross-correlations, the determinations of inference from data streams are associated with their \textit{optimal} signal-to-noise ratio (SNR) \cite{Allen:2005fk},
\begin{equation} \label{eq:snr}
\begin{aligned}
    \rho^2_\mathrm{TDI} = & 4  \int^{\infty}_0 \frac{ | \tilde{h}_\mathrm{TDI} (f) |^2 }{ S_\mathrm{TDI} (f) }  d f \\
    = & \int^{\infty}_0 \left( \frac{ 2 | \tilde{h}_\mathrm{TDI} (f) | \sqrt{f} }{ \sqrt{S_\mathrm{TDI} (f) } } \right)^2  d \ln f .
\end{aligned}
\end{equation}
By employing the waveform of source2, the values of $\frac{ 2 | \tilde{h}_\mathrm{TDI} (f) | \sqrt{f} }{ \sqrt{S_\mathrm{TDI} (f) } }$ for the optimal TDI channels are shown in Fig. \ref{fig:delta_snr}. The upper left panel shows the curves for (A$_\mathrm{U\bar{U}}$, E$_\mathrm{U\bar{U}}$, T$_\mathrm{U\bar{U}}$), and the upper right plot presents the results for (A$_\mathrm{X1}$, E$_\mathrm{X1}$, T$_\mathrm{X1}$). Grey vertical lines mark the CFs of each TDI configuration. 
The curves illustrate the science channels, A and E, dominate the SNR in the lower frequencies. The T$_\mathrm{U\bar{U}}$ makes a trivial contribution in the low frequency band, while the T$_\mathrm{X1}$ exhibits a moderate SNR contribution in the same range. Our previous analysis showed that this gain is produced by the difference between the ordinary channels (X1, Y1, Z1). In other words, the signal in T$_\mathrm{X1}$ arises from the imperfect cancellation of X1, Y1, and Z1 \cite{Wang:2020fwa}, rendering it susceptible to variations in interferometer arms over time. And this residual signal in T channel is tricky to be accurately modeled in frequency-domain. Consequently, inaccurate signal subtraction in the T$_\mathrm{X1}$ channel could introduce bias into the analysis.
In this analysis, PSDs are estimated by averaging data over 30 days, which may also leading to inaccurate estimations in T$_\mathrm{X1}$. As a result, including the T$_\mathrm{X1}$ data stream biased the inference results. 

\begin{figure}[htb]
\includegraphics[width=0.48\textwidth]{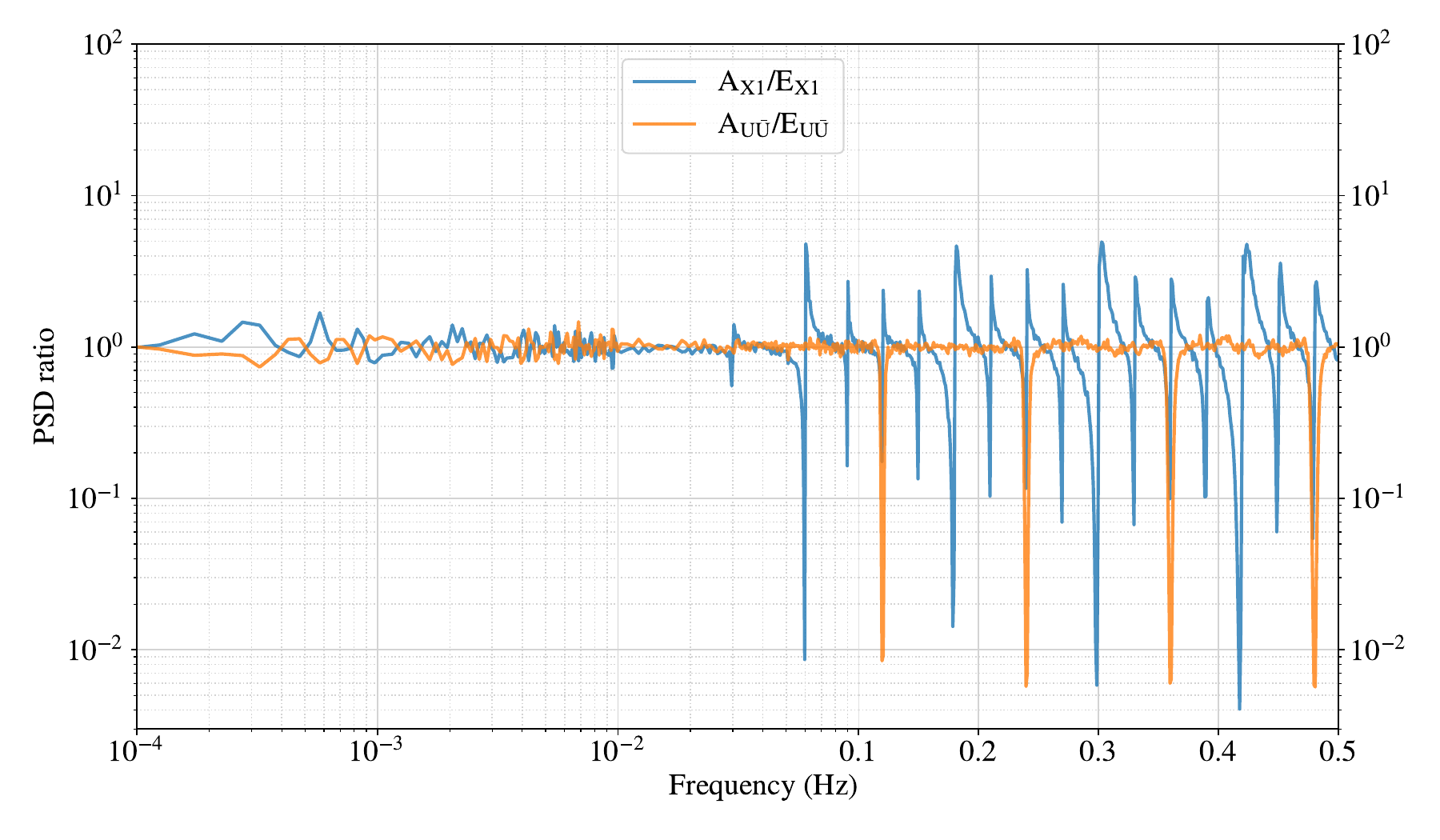}
\caption{\label{fig:psd_ratio_X1_UUbar} The PSD ratios of A and E channels for Michelson (blue) and hybrid Relay (orange) configurations. The discontinuities in the ratio of Michelson indicate that the CFs of A$_\mathrm{X1}$ and E$_\mathrm{X1}$ are slightly different resulting from the different unequal arms involved in the channels. The PSDs from A$_\mathrm{U\bar{U}}$ and E$_\mathrm{U\bar{U}}$ are more identical and stable in the sensitive band except at their CFs.}
\end{figure}

In the higher frequency band, the ratios shapely drop at the CFs due to significant suppression of the GW signal and the overestimation of PSD at the CFs caused by smoothing of averaging. The hybrid Relay family is subject to minimal CFs, and the second-generation Michelson suffers from more CFs. To compare the joint SNR from three observables, the square root of quadratic sums over A, E, and T, $\sqrt{ \sum_\mathrm{A, E, T} \frac{4 |\tilde{h}(f)|^2 f }{S} }$, are calculated and shown in the lower panel. The curve of hybrid Relay configuration is smoother and more robust, whereas the Michelson observables may slightly lose SNR around the CFs.

For the Michelson channels, another potential problem is the slight difference in CFs between the science channels, A and E. The PSD ratios of A and E, $S_A/S_E$, for Michelson and hybrid Relay, are depicted in Fig. \ref{fig:psd_ratio_X1_UUbar}. Below 50 mHz, the PSD of A$_\mathrm{X1}$ and E$_\mathrm{X1}$ are closely equal, but as the frequency increase, their PSDs diverge especially around their CFs. In this data chunk, the CF of A$_\mathrm{X1}$ is slightly lower than E$_\mathrm{X1}$ which causes the ratio jumps from low to high at each CF. The rhythm of CFs in the hybrid Relay is more stable, the two channels have the same CFs. The ratios are dropped at each CF due to the PSD of A$_\mathrm{U\bar{U}}$ is lower than the E$_\mathrm{U\bar{U}}$'s in current case. The asynchronous CFs in Michelson result from the different unequal arms involved in the channels. Optimal observables from the hybrid Relay have the same CFs, potentially enhancing the stability of their covariance matrix and thus benefiting the robustness of data analysis.

\section{Conclusion and discussion} \label{sec:conclusions}

In this work, we present a hybrid Relay second-generation TDI configuration with minimal characteristic frequencies. We compare its capability to the fiducial second-generation TDI Michelson configuration, focusing on laser noise suppression, clock noise cancellation, and GW signal inference as crucial aspects. To evaluate their performance in data analysis, we simulate and analyze data containing GW signals from massive binary black holes. 
The results demonstrate that new TDI observables effectively satisfy the requirements for laser noise and clock noise suppressions as the Michelson configuration. The hybrid Relay configuration also exhibits robust performance in parameter inference, with noise spectra and GW response function less modulated benefiting from fewer characteristic frequencies. In contrast, the TDI observables from the Michelson configuration would suffer from unequal arm lengths during orbit motion, particularly affecting the T channel at low frequencies which undermines parameter inference. Additionally, the Michelson observables will suffer from more characteristic frequencies where GW signals and noises are significantly suppressed. On the other side, other TDI configurations with asymmetric geometries could also have minimal characteristic frequencies, for instance, the PD observable from mixed Beacon and Monitor as described in \cite{Wang:2020pkk}. The hybrid Relay should be one of the most straightforward options to implement and readily compatible with current noise solutions of the fiducial Michelson configuration.

We specifically focus on GW signals from massive binary black holes to illustrate our analysis. The science data streams, A and E, dominate the inference outcome, and the T channel is insignificant and negligible in the analysis for such sources.
However, the T observable, being noise-dominant, is crucial for characterizing instrumental noises and separating stochastic GW \cite{Wang:2022sti,Muratore:2022nbh,Hartwig:2023pft}. The T data from Michelson is susceptible to time-varying arms and contamination by GW signal, rendering it ineligible for this task. The T channel from the hybrid Relay, being genuinely noise-dominating data, could be more suitable for noise characterization.
On the other side, the selected chirp signal quickly sweeps over the high frequency band which weakens the impacts of characteristic frequencies.
For a slowly evolved GW signal, such as the early inspiral of stellar mass binaries, the signal will last much longer around TDI's characteristic frequencies, and then the signal will be dramatically suppressed and modulated. Observables with fewer null frequencies will promote signal recoveries by reducing modulations and suppressions.

Machine learning techniques have been employed to boost the efficiency of GW identifications and estimations \cite[and reference therein]{Williams:2021qyt,Gabbard:2019rde,Langendorff:2022fzq,Dax:2022pxd,Ruan:2021fxq,Ruan:2023fce,Du:2023plr,Sun:2023vlq,Xu:2024jbo}. The concept of global fits has been proposed to simultaneously resolve various GW signals \cite{Littenberg:2020bxy,Littenberg:2023xpl}. TDI observables with stable PSDs and GW response functions are essential for these analyses. The proposed hybrid Relay could be the eligible configuration to fulfill these requirements, and we are committed to further studies in future work.

\begin{acknowledgments}

GW was supported by the National Key R\&D Program of China under Grant No. 2021YFC2201903, and NSFC No. 12003059. This work made use of the High Performance Computing Resource in the Core Facility for Advanced Research Computing at Shanghai Astronomical Observatory.
This work are performed by using the python packages $\mathsf{numpy}$ \cite{harris2020array}, $\mathsf{scipy}$ \cite{2020SciPy-NMeth}, $\mathsf{pandas}$ \cite{pandas}, PyCBC \cite{Usman:2015kfa,Biwer:2018osg}, LALSuite \cite{lalsuite,swiglal}, $\mathsf{astropy}$ \cite{Astropy:2013muo}, $\mathsf{BILBY}$ \cite{Ashton:2018jfp}, $\mathsf{MultiNest}$ \cite{Feroz:2008xx} and $\mathsf{PyMultiNest}$ \cite{Buchner:2014nha}, and the plots are make by utilizing $\mathsf{matplotlib}$ \cite{Hunter:2007ouj}, $\mathsf{GetDist}$ \cite{Lewis:2019xzd}. 

\end{acknowledgments}

\appendix

\section{PSD and CSD of hybrid Relay TDI observables} \label{sec:psd_csd}

\begin{figure*}[htb]
\includegraphics[width=0.48\textwidth]{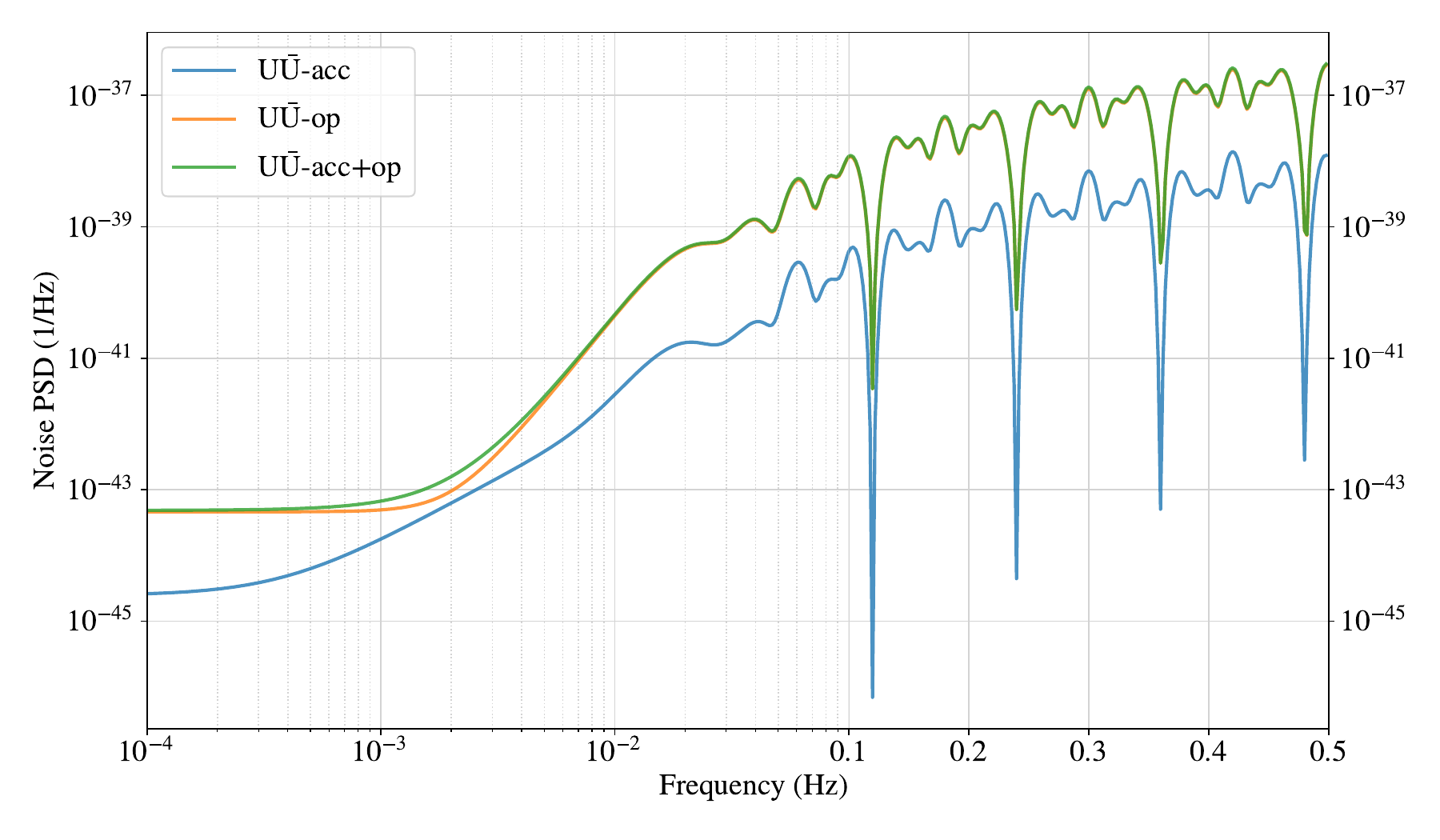}
\includegraphics[width=0.48\textwidth]{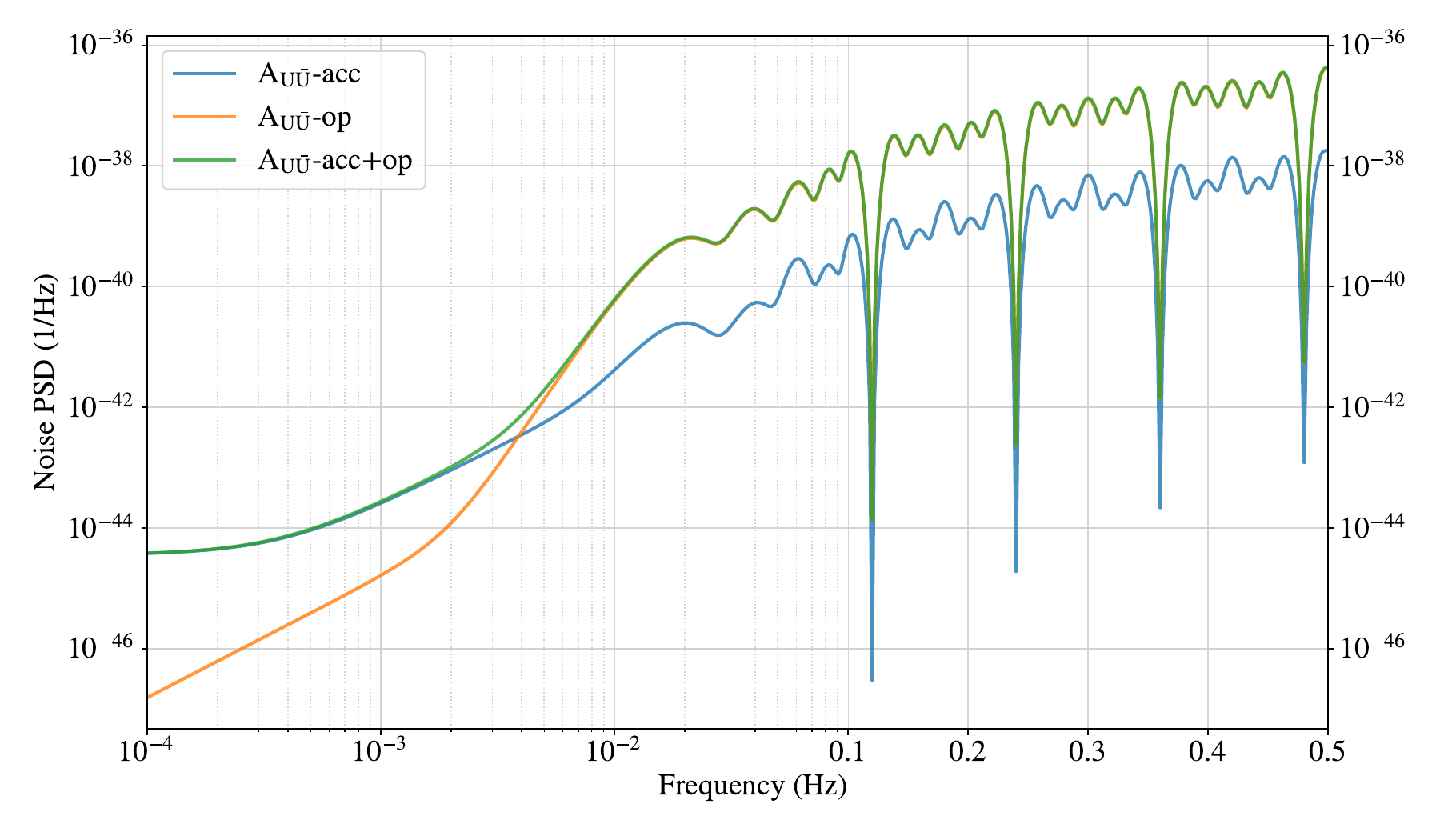}
\includegraphics[width=0.48\textwidth]{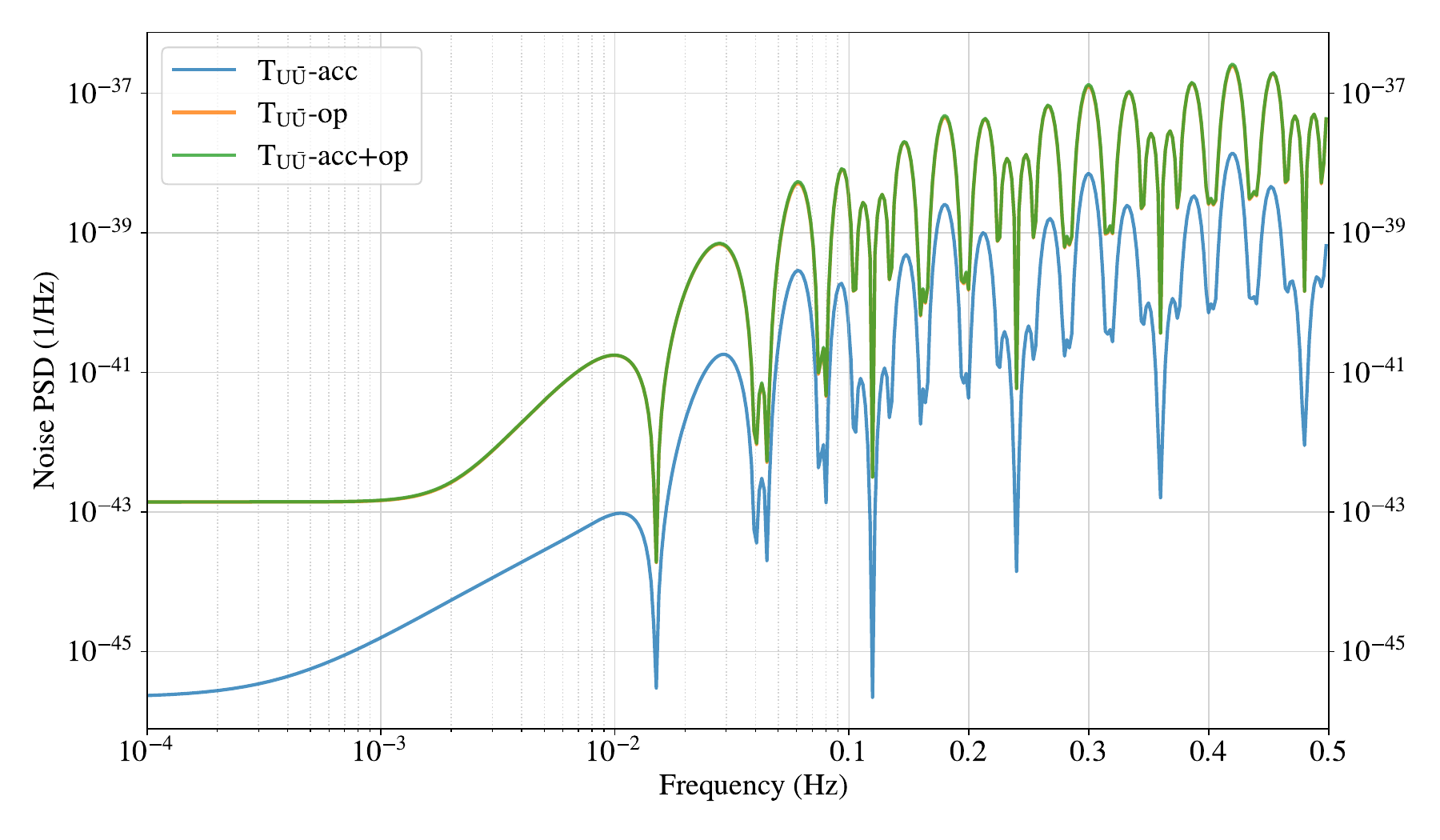}
\caption{\label{fig:psd_UUbar_inst} The PSDs of $\mathrm{U\bar{U}}$, A$_\mathrm{U\bar{U}}$, and T$_\mathrm{U\bar{U}}$ observables with separating acceleration noises (acc) and optical metrology noises (op). In the $\mathrm{U\bar{U}}$ channel, optical metrology noise dominates over acceleration except within the range of [1, 4] mHz. The acceleration noise is higher than the optical path noise in low frequencies for $A_\mathrm{U\bar{U}}$, and optical noise becomes the dominant noise in higher frequencies. In the $T_\mathrm{U\bar{U}}$ channel, the optical metrology noise overwhelms acceleration noise across the entire frequency band.}
\end{figure*}

The PSDs of secondary noises separating the acceleration noise and optical metrology noise are depicted in Fig. \ref{fig:psd_UUbar_inst}. The coefficients of acceleration noises and optical metrology noises for both PSD and CSD are provided in Tables \ref{tab:TDI_PSD} and \ref{tab:TDI_CSD}, respectively.
In the $\mathrm{U\bar{U}}$ channel, optical metrology noise predominates over acceleration except within the frequency range of [1, 4] mHz. In the A$_\mathrm{U\bar{U}}$ channels, acceleration noise surpasses optical path noise in the lower frequency band, while optical noise becomes a dominant factor at higher frequencies. In the T$_\mathrm{U\bar{U}}$ channel, optical metrology noise overwhelms acceleration noise across the entire frequency band.

\begin{table*}[htb]
\caption{\label{tab:TDI_PSD} The coefficients of noise components for PSD (power-spectral density) of $\mathrm{U\bar{U}}$ and their optimal TDI channels ($x = 2 \pi f L$). }
\begin{ruledtabular}
\begin{tabular}{c|c|c}
 & $\mathrm{U\bar{U}}$ & A$_\mathrm{U\bar{U}}$ \\
\hline
 $S_{\mathrm{acc}12 }$ & $ 64 \cos^2 x \sin^4 x $  & $32 \cos^2 x (5 + 6 \cos x)^2 \sin^4 \frac{x}{2}$  \\
  $S_{\mathrm{acc}21 }$ & $ 16 \sin^4 x $ & $32 \sin ^4 \frac{x}{2} (5 \cos x+2 \cos 2x+\cos 3x+3)^2$  \\
 $S_{\mathrm{acc}13 }$ & $ 64 \cos^2 x \sin^4 x $ & $8 \sin ^4 x (1-2 \cos x)^2$ \\
 $S_{\mathrm{acc}31 }$ & $ 16 (1 + 2 \cos 2 x )^2 \sin^4 x $ & $8 \sin ^4x (-2 \cos x+2 \cos 2x+1)^2$ \\
 $S_{\mathrm{acc}23 }$ & $ 64 \cos^2 x (3 + 4 \cos x)^2 \sin^4 \frac{x}{2} $ & $32 \sin ^4 \frac{x}{2} (5 \cos x+2 \cos 2x+\cos 3x+3)^2$ \\
 $S_{\mathrm{acc}32 }$ & $ 64 \cos^2 x \sin^4 \frac{3 x}{2} $ & $32 \sin ^4 \frac{x}{2} (6 \cos x+3 \cos 2x+\cos 3x+3)^2$ \\
\hline
 $S_{\mathrm{op}12 }$ & $ 4 \sin^2 x $  & $8 \sin ^4 \frac{x}{2} (46 \cos x+30 \cos 2x+14 \cos 3x+4 \cos 4x+27)$ \\
  $S_{\mathrm{op}21 }$ & $ 4 \sin^2 x $ & $8 \sin ^4 \frac{x}{2} (46 \cos x+30 \cos 2x+14 \cos 3x+4 \cos 4x+27)$ \\
 $S_{\mathrm{op}13 }$ & $ 4 \sin^2 x $ & $8 \sin ^2 \frac{x}{2} \sin ^2x$  \\
  $S_{\mathrm{op}31 }$ & $ 4 \sin^2 x $ & $8 \sin ^2 \frac{x}{2} \sin ^2x$ \\
 $S_{\mathrm{op}23 }$ & $2 (2 - \cos x + \cos 2x - 2 \cos 3x - \cos 5x + \cos 6x )$ & $8 \sin ^4 \frac{x}{2} (58 \cos x+42 \cos 2x+24 \cos 3x+10 \cos 4x+2 \cos 5x+33)$ \\
 $S_{\mathrm{op}32 }$ & $2 (2 - \cos x + \cos 2x - 2 \cos 3x - \cos 5x + \cos 6x )$ & $8 \sin^4 \frac{x}{2} (58 \cos x+42 \cos 2x+24 \cos 3x+10 \cos 4x+2 \cos 5x+33)$  \\
 \hline \hline
 & T$_\mathrm{U\bar{U}}$ &  E$_\mathrm{U\bar{U}}$  \\
 \hline
 $S_{\mathrm{acc}12 }$ & $ \frac{64}{3} \sin ^4 \frac{x}{2}  (\cos x+\cos 2x+\cos 3x)^2 $ & $ \frac{32}{3} \sin ^4 \frac{x}{2}  (5 \cos x+5 \cos 2x+2 \cos 3x+3)^2 $   \\
    $S_{\mathrm{acc}21 }$ & $ \frac{64}{3} \sin ^4 \frac{x}{2}  (\cos x+\cos 2x+\cos 3x)^2 $ & $ \frac{32}{3} \sin ^4 \frac{x}{2}  (7 \cos x+4 \cos 2x+\cos 3x+3)^2 $  \\
 $S_{\mathrm{acc}13 }$ & $ \frac{64}{3} \sin ^4 \frac{x}{2}  (\cos x+\cos 2x+\cos 3x)^2 $ & $ \frac{32}{3} \sin ^4 \frac{x}{2}  (11 \cos x+5 \cos 2x+2 \cos 3x+6)^2$  \\
  $S_{\mathrm{acc}31 }$ & $ \frac{64}{3} \sin ^4 \frac{x}{2}  (\cos x+\cos 2x+\cos 3x)^2 $ & $ \frac{32}{3} \sin ^4 \frac{x}{2}  (10 \cos x+7 \cos 2x+\cos 3x+6)^2$  \\
 $S_{\mathrm{acc}23 }$ & $ \frac{64}{3} \sin ^4 \frac{x}{2}  (\cos x+\cos 2x+\cos 3x)^2 $ & $\frac{32}{3} \sin ^4 \frac{x}{2}  (5 \cos x+2 \cos 2x-\cos 3x+3)^2$  \\
 $S_{\mathrm{acc}32 }$ & $ \frac{64}{3} \sin ^4 \frac{x}{2}  (\cos x+\cos 2x+\cos 3x)^2 $ & $ \frac{32}{3} \sin ^4 \frac{x}{2}  (4 \cos x+\cos 2x+\cos 3x+3)^2$  \\
\hline
 $S_{\mathrm{op}12 }$ & $ \frac{4}{3} \left(\sin  \frac{x}{2} -\sin  \frac{7 x}{2} \right)^2 $ & $ \frac{8}{3} \sin ^4 \frac{x}{2} (74 \cos x+54 \cos 2 x+34 \cos 3x+16 \cos 4x+4 \cos 5x+43) $   \\
  $S_{\mathrm{op}21 }$ & $ \frac{4}{3} \left(\sin  \frac{x}{2} -\sin  \frac{7 x}{2} \right)^2 $ & $ \frac{8}{3} \sin ^4 \frac{x}{2}  (74 \cos x+54 \cos 2x+34 \cos 3x+16 \cos 4x+4 \cos 5x+43) $  \\
 $S_{\mathrm{op}13 }$ & $ \frac{4}{3} \left(\sin  \frac{x}{2} -\sin  \frac{7 x}{2} \right)^2 $ & $ \frac{16}{3} \sin ^4 \frac{x}{2}  (103 \cos x+72 \cos 2x+38 \cos 3x+14 \cos 4x+2 \cos 5x+59) $  \\
 $S_{\mathrm{op}31 }$ & $ \frac{4}{3} \left(\sin  \frac{x}{2} -\sin  \frac{7 x}{2} \right)^2 $ & $ \frac{16}{3} \sin ^4 \frac{x}{2}  (103 \cos x+72 \cos 2x+38 \cos 3x+14 \cos 4x+2 \cos 5 x+59) $  \\
 $S_{\mathrm{op}23 }$ & $ \frac{4}{3} \left(\sin  \frac{x}{2} -\sin  \frac{7 x}{2} \right)^2 $ & $ \frac{8}{3} \sin ^4 \frac{x}{2}  (38 \cos x+18 \cos 2x+4 \cos 3x-2 \cos 4x-2 \cos 5x+25) $  \\
 $S_{\mathrm{op}32 }$ & $ \frac{4}{3} \left(\sin  \frac{x}{2} -\sin  \frac{7 x}{2} \right)^2 $ & $ \frac{8}{3} \sin ^4 \frac{x}{2}  (38 \cos x+18 \cos 2x+4 \cos 3x-2 \cos 4x-2 \cos 5x+25) $  \\
 \end{tabular}
\end{ruledtabular}
\end{table*}

\begin{table*}[htb]
\caption{\label{tab:TDI_CSD} The coefficients of noise components for CSD (cross-spectral density) of $\mathrm{U\bar{U}}$ and their optimal TDI channels ($x = 2 \pi f L$). }
\begin{ruledtabular}
\fontsize{8pt}{8pt}\selectfont
\centering
\begin{tabular}{c|c}
 & CSD($\mathrm{U\bar{U}},\mathrm{V\bar{V}}$)  \\
\hline
$S_{\mathrm{acc}12 }$ & $32 \sin ^4x (2 \cos x+\cos 3x) $   \\
$S_{\mathrm{acc}21 }$ & $32 \sin ^4x \cos x$   \\
$S_{\mathrm{acc}13 }$ &  $-64 \sin ^4 \frac{x}{2}  \left(3 \cos  \frac{x}{2} +2 \cos \frac{3 x}{2} + \cos \frac{5 x}{2} \right)^2$   \\
$S_{\mathrm{acc}31 }$ & $2 (-\cos 2x+\cos 4x+2 \cos 5x+\cos 6x-2 \cos 7x-1)$   \\
$S_{\mathrm{acc}23 }$ & $ -64 \sin ^4 \frac{x}{2}  (4 \cos x+3) \left(\cos  \frac{x}{2} +\cos \frac{3 x}{2} \right)^2$  \\
$S_{\mathrm{acc}32 }$ & $-128 \sin ^4 \frac{x}{2}  \cos x \left(2 \cos  \frac{x}{2} +\cos \frac{3 x}{2} \right)^2$   \\
$S_{\mathrm{op}12 }$ & $4 \sin ^2x \cos x + 4 i \sin ^3x$  \\
$S_{\mathrm{op}21 }$ & $ 4 \sin ^2x \cos x - 4 i \sin ^3x $  \\
$S_{\mathrm{op}13 }$ & $ (\cos 4x+\cos 5x-\cos 7x-1) -  i (2 \sin x-\sin 4x-\sin 5x+\sin 7x) $  \\
$S_{\mathrm{op}31 }$ & $ (\cos 4x+\cos 5x-\cos 7x-1) + i (2 \sin x-\sin 4x-\sin 5x+\sin 7x) $   \\
$S_{\mathrm{op}23 }$ & $ 4 \sin ^2x (-\cos x+\cos 2x+\cos 4x)  + i (\sin x-\sin 2x+\sin 3x+\sin 4x-\sin 6x) $  \\
$S_{\mathrm{op}32 }$ & $ 4 \sin ^2x (-\cos x+\cos 2x+\cos 4x) - i (\sin x-\sin 2x+\sin 3x+\sin 4x-\sin 6x) $   \\
\hline
& CSD(A$_\mathrm{U\bar{U}}$,E$_\mathrm{U\bar{U}}$ ) \\
\hline
$S_{\mathrm{acc}12 }$ &  $\frac{16}{\sqrt{3}} \sin ^4 \frac{x}{2}  (106 \cos x+83 \cos 2x+52 \cos 3x+25 \cos 4x+6 \cos 5x+58)$  \\
$S_{\mathrm{acc}21 }$ & $ \frac{16}{\sqrt{3}}  \sin ^4 \frac{x}{2}  (112 \cos x+83 \cos 2x+46 \cos 3x+20 \cos 4x+6 \cos 5x+\cos 6x+62) $  \\
$S_{\mathrm{acc}13 }$ &  $ -\frac{64}{\sqrt{3}}  \sin ^4 \frac{x}{2}  \cos ^2 \frac{x}{2}  (6 \cos x+8 \cos 2x+3 \cos 3x+2 \cos 4x+5)$  \\
$S_{\mathrm{acc}31 }$ & $ -\frac{64 }{\sqrt{3}}  \sin ^4 \frac{x}{2}  \cos ^2 \frac{x}{2}  (2 \cos x+8 \cos 2x+4 \cos 3x+6 \cos 4x+\cos 5x+3)$  \\
$S_{\mathrm{acc}23 }$ & $ \frac{16}{\sqrt{3}}  \sin ^4 \frac{x}{2}  (-88 \cos x-61 \cos 2x-30 \cos 3x-8 \cos 4x+2 \cos 5x+\cos 6x-50) $  \\
$S_{\mathrm{acc}32 }$ & $ -\frac{16}{\sqrt{3}} \sin ^4 \frac{x}{2}  (82 \cos x+58 \cos 2x+30 \cos 3x+13 \cos 4x+4 \cos 5x+\cos 6x+46) $  \\
$S_{\mathrm{op}12 }$ &  $ \frac{8}{\sqrt{3}}  \sin ^4 \frac{x}{2}  (56 \cos x+42 \cos 2x+24 \cos 3x+10 \cos 4x+2 \cos 5x+31)+ \frac{i}{\sqrt{3}} (6 \sin x-\sin 2x-\sin 5x-\sin 6x+\sin 7x) $   \\
$S_{\mathrm{op}21 }$ &  $ \frac{8}{\sqrt{3}}  \sin ^4 \frac{x}{2}  (56 \cos x+42 \cos 2x+24 \cos 3x+10 \cos 4x+2 \cos 5x +31) - \frac{i}{\sqrt{3}}  (6 \sin x-\sin 2x - \sin 5x - \sin 6x + \sin 7x) $  \\
$S_{\mathrm{op}13 }$ & $ -\frac{64}{\sqrt{3}} \sin ^4 \frac{x}{2}  \cos ^2 \frac{x}{2}  (\cos x+2 \cos 2x+\cos 3x+\cos 4x+1) - \frac{i}{\sqrt{3}} (6 \sin x-\sin 2x-\sin 5x-\sin 6x+\sin 7x)  $   \\
$S_{\mathrm{op}31 }$ & $ -\frac{64}{\sqrt{3}} \sin ^4 \frac{x}{2}  \cos ^2 \frac{x}{2}  (\cos x+2 \cos 2x+\cos 3x+\cos 4x+1) + \frac{i}{\sqrt{3}} (6 \sin x-\sin 2x-\sin 5x-\sin 6x+\sin 7x) $   \\
$S_{\mathrm{op}23 }$ & $ -\frac{8}{\sqrt{3}} \sin ^4 \frac{x}{2}  (2 \cos x+1)^2 (6 \cos x+4 \cos 2x+3)  + \frac{i}{\sqrt{3}}  (6 \sin x-\sin 2x-\sin 5x-\sin 6x+\sin 7x)$   \\
$S_{\mathrm{op}32 }$ & $ -\frac{8}{\sqrt{3}}  \sin ^4 \frac{x}{2}  (2 \cos x+1)^2 (6 \cos x+4 \cos 2x+3) - \frac{i}{\sqrt{3}}  (6 \sin x-\sin 2x-\sin 5x-\sin 6x+\sin 7x)$   \\
\hline
 & CSD(A$_\mathrm{U\bar{U}}$,T$_\mathrm{U\bar{U}}$ )\\
\hline
$S_{\mathrm{acc}12 }$ &  $ -16 \sqrt{\frac{2}{3}} \sin ^4 \frac{x}{2}  (17 \cos x+16 \cos 2x+14 \cos  3x +8 \cos  4x +3 \cos  5x +8) $  \\
$S_{\mathrm{acc}21 }$ & $ 16 \sqrt{\frac{2}{3}} \sin ^4 \frac{x}{2}  (16 \cos x+17 \cos 2x+13 \cos  3x +8 \cos  4x +3 \cos  5x +\cos  6x +8) $   \\
$S_{\mathrm{acc}13 }$ & $64 \sqrt{\frac{2}{3}} \sin ^4 \frac{x}{2}  \cos ^2 \frac{x}{2}  (\cos 2x+\cos  4x +1) $   \\
$S_{\mathrm{acc}31 }$ &  $ -64 \sqrt{\frac{2}{3}} \sin ^4 \frac{x}{2}  \cos ^2 \frac{x}{2}  (2 \cos x-\cos 2x+\cos  3x +\cos  5x ) $  \\
$S_{\mathrm{acc}23 }$ &  $ 16 \sqrt{\frac{2}{3}} \sin ^4 \frac{x}{2}  (20 \cos x+17 \cos 2x+15 \cos  3x +10 \cos  4x +5 \cos  5x +\cos  6x +10) $  \\
$S_{\mathrm{acc}32 }$ &  $ -16 \sqrt{\frac{2}{3}} \sin ^4 \frac{x}{2}  (19 \cos x+19 \cos 2x+15 \cos  3x +10 \cos  4x +4 \cos  5x +\cos  6x +10) $  \\
$S_{\mathrm{op}12 }$ & $-8 \sqrt{\frac{2}{3}} \sin ^4 \frac{x}{2}  (4 \cos x+3 \cos 2x+3 \cos  3x +2 \cos  4x +\cos  5x +2) - \frac{i}{\sqrt{6}}  (3 \sin x+2 \sin 2x-3 \sin  3x -3 \sin  4x -\sin  5x +2 \sin  6x +\sin  7x ) $    \\
$S_{\mathrm{op}21 }$ &  $ -8 \sqrt{\frac{2}{3}} \sin ^4 \frac{x}{2}  (4 \cos x+3 \cos 2x+3 \cos  3x +2 \cos  4x +\cos  5x +2)  + \frac{i}{\sqrt{6}} (3 \sin x+2 \sin 2x-3 \sin  3x -3 \sin  4x -\sin  5x +2 \sin  6x +\sin  7x ) $  \\
$S_{\mathrm{op}13 }$ &  $ 32 \sqrt{\frac{2}{3}} \sin ^4 \frac{x}{2}  \cos ^2 \frac{x}{2}  (\cos x+2 \cos 2x+\cos  3x +\cos  4x +1) - 16 i \sqrt{\frac{2}{3}} \sin ^3 \frac{x}{2}  \cos ^2 2 x  \left(2 \cos \frac{x}{2} + \cos \frac{3 x}{2} \right) $   \\
$S_{\mathrm{op}31 }$ &  $ 32 \sqrt{\frac{2}{3}} \sin ^4 \frac{x}{2}  \cos ^2 \frac{x}{2}  (\cos x+2 \cos 2x+\cos  3x +\cos  4x +1) + 16 i \sqrt{\frac{2}{3}} \sin ^3 \frac{x}{2}  \cos ^2 2 x  \left(2 \cos \frac{x}{2} + \cos \frac{3 x}{2} \right) $   \\
$S_{\mathrm{op}23 }$ &  $-8 \sqrt{\frac{2}{3}} \sin ^4 \frac{x}{2}  (2 \cos x+1)^2 \cos 2x + \frac{i}{\sqrt{6}} (\sin 2x-3 \sin  4x -2 \sin  5x +\sin  6x +2 \sin  7x )  $   \\
$S_{\mathrm{op}32 }$ &  $ -8 \sqrt{\frac{2}{3}} \sin ^4 \frac{x}{2}  (2 \cos x+1)^2 \cos 2x -\frac{i}{\sqrt{6}}  (\sin 2x-3 \sin  4x -2 \sin  5x +\sin  6x +2 \sin  7x ) $   \\
\hline
& CSD( E$_\mathrm{U\bar{U}}$,T$_\mathrm{U\bar{U}}$ ) \\
\hline
$S_{\mathrm{acc}12 }$ &  $ -\frac{16}{3} \sqrt{2} \sin ^4 \frac{x}{2}  (2 \cos x+1)^2 (3 \cos x+3 \cos  3x +2 \cos  4x +2) $   \\
$S_{\mathrm{acc}21 }$ & $ \frac{16}{3} \sqrt{2} \sin ^4 \frac{x}{2}  (22 \cos x+21 \cos 2x+17 \cos  3x +12 \cos  4x +5 \cos  5x +\cos  6x +12) $   \\
$S_{\mathrm{acc}13 }$ &  $ -\frac{16}{3} \sqrt{2} \sin ^4 \frac{x}{2}  (2 \cos x+1)^2 (3 \cos x+6 \cos 2x+3 \cos  3x +2 \cos  4x +2) $   \\
$S_{\mathrm{acc}31 }$ &  $ \frac{16}{3} \sqrt{2} \sin ^4 \frac{x}{2}  (37 \cos x+33 \cos 2x+29 \cos  3x +18 \cos  4x +8 \cos  5x +\cos  6x +18) $  \\
$S_{\mathrm{acc}23 }$ &  $ -\frac{16}{3} \sqrt{2} \sin ^4 \frac{x}{2}  (14 \cos x+15 \cos 2x+13 \cos  3x +6 \cos  4x +\cos  5x -\cos  6x +6) $  \\
$S_{\mathrm{acc}32 }$ & $ \frac{16}{3} \sqrt{2} \sin ^4 \frac{x}{2}  (13 \cos x+15 \cos 2x+11 \cos  3x +6 \cos  4x +2 \cos  5x +\cos  6x +6) $   \\
$S_{\mathrm{op}12 }$ & $ \frac{8}{3} \sqrt{2} \sin ^4 \frac{x}{2}  (8 \cos x+9 \cos 2x+7 \cos  3x +4 \cos  4x +\cos  5x +4) + \frac{i}{\sqrt{2}} (\sin x-\sin  3x +\sin  4x +\sin  5x -\sin  7x ) $   \\
$S_{\mathrm{op}21 }$ & $ \frac{8}{3} \sqrt{2} \sin ^4 \frac{x}{2}  (8 \cos x+9 \cos 2x+7 \cos  3x +4 \cos  4x +\cos  5x +4) -\frac{i}{\sqrt{2}} (\sin x-\sin  3x +\sin  4x +\sin  5x -\sin  7x )  $  \\
$S_{\mathrm{op}13 }$ &  $ \frac{16}{3} \sqrt{2} \sin ^4 \frac{x}{2}  (2 \cos x+1) \cos ^2 2x + 4 i \sqrt{2} \sin ^3x (\cos x+2 \cos 2x+\cos  3x +\cos  4x +1) $   \\
$S_{\mathrm{op}31 }$ & $ \frac{16}{3} \sqrt{2} \sin ^4 \frac{x}{2}  (2 \cos x+1) \cos ^22x -4 i \sqrt{2} \sin ^3x (\cos x+2 \cos 2x+\cos  3x +\cos  4x +1) $   \\
$S_{\mathrm{op}23 }$ & $ -\frac{8}{3} \sqrt{2} \sin ^4 \frac{x}{2}  (10 \cos x+9 \cos 2x+8 \cos  3x +5 \cos  4x +2 \cos  5x +5)  + 8 i \sqrt{2} \sin ^3 \frac{x}{2}  \cos  \frac{x}{2}  (2 \cos x+1)^2 \cos 2x $  \\
$S_{\mathrm{op}32 }$ &  $ -\frac{8}{3} \sqrt{2} \sin ^4 \frac{x}{2}  (10 \cos x+9 \cos 2x+8 \cos  3x +5 \cos  4x +2 \cos  5x +5) -8 i \sqrt{2} \sin ^3 \frac{x}{2}  \cos  \frac{x}{2}  (2 \cos x+1)^2 \cos 2x $  \\
 \end{tabular}
\end{ruledtabular}
\end{table*}

\nocite{*}
\bibliography{apsref}

\end{document}